%% file: main.tex
\documentclass[reprint,superscriptaddress,aps,pra,floatfix,nofootinbib]{revtex4-2}

\input{packages}

\input{commands}

\begin{document}

\preprint{APS/123-QED}
\title{Scalable Hessian-vector products for tensor networks with application to quantum circuit compression}

\author{Isabel~Nha~Minh~Le\orcidlink{0000-0001-6707-044X}}
\email{isabel.le@tum.de}
\affiliation{Technical University of Munich, School of Computation, Information and Technology, Department of Computer Science, Boltzmannstraße 3, 85748 Garching, Germany}
\affiliation{Munich Center for Quantum Science and Technology (MCQST), Schellingstrasse 4, 80799 Munich, Germany}
\author{Roeland~Wiersema\orcidlink{0000-0002-0839-4265}}
\email{rwiersema@flatironinstitute.org}
\affiliation{Center for Computational Quantum Physics, Flatiron Institute, 162 Fifth Avenue, New York, NY 10010, USA}
\author{Christian~B.~Mendl\orcidlink{0000-0002-6386-0230}} 
\email{christian.mendl@tum.de}
\affiliation{Technical University of Munich, School of Computation, Information and Technology, Department of Computer Science, Boltzmannstraße 3, 85748 Garching, Germany}
\affiliation{Munich Center for Quantum Science and Technology (MCQST), Schellingstrasse 4, 80799 Munich, Germany}
\affiliation{Technical University of Munich, Institute for Advanced Study, Lichtenbergstraße 2a, 85748 Garching, Germany}

\date{\today}

\input{0_abstract}

\maketitle

\input{1_introduction}
\input{2_general-method}
\input{3_quantum-setting}
\input{5_application}
\input{6_conclusion}
\input{7_acknowledgement}

\input{acronyms}
\bibliography{bibliography.bib}

\clearpage
\appendix
\input{8_appendix}

\end{document}

%% file: packages.tex
\pdfoutput=1
\usepackage[main=english]{babel}
\usepackage[dvipsnames]{xcolor}
\usepackage{colortbl}
\usepackage{braket}
\usepackage{array}
\usepackage{amsmath,amssymb,amsfonts,amsthm}
\usepackage[export]{adjustbox}
\usepackage{graphicx,wrapfig,lipsum}
\usepackage{mathtools}
\usepackage{caption}
\usepackage{subcaption}
\usepackage[colorlinks=true,linkcolor=magenta,urlcolor=magenta,citecolor=magenta,breaklinks=true]{hyperref}
\usepackage[capitalise]{cleveref}
\usepackage{physics}
\usepackage{float}
\usepackage{multirow}
\usepackage[utf8]{inputenc}
\usepackage[toc,page]{appendix}
\usepackage{dsfont}
\usepackage{xspace}
\usepackage{suffix}
\usepackage{xstring}
\usepackage[nolist, nohyperlinks]{acronym}
\usepackage{comment}
\usepackage{adjustbox}
\usepackage{enumerate}
\usepackage{listings}
\usepackage{algorithm}
\usepackage{algpseudocode}
\usepackage{listings}
\usepackage{bm}
\usepackage{orcidlink}
\usepackage{booktabs}

%% file: commands.tex
\renewcommand{\tangent}{tangent\xspace}

\newcommand{\mc}[1]{\ensuremath{\mathcal{#1}}}
\newcommand{\pa}[1]{\left( #1 \right)}    
\newcommand{\br}[1]{\left[ #1 \right]}    
\newcommand{\cb}[1]{\left\{ #1 \right\}}  
\renewcommand{\abs}[1]{\left| #1 \right|}   

\newcommand{\unitary}{\ensuremath{\mathcal{U}}}

\newcommand{\C}{\ensuremath{\mathbb{C}}}
\newcommand{\R}{\ensuremath{\mathbb{R}}}
\newcommand{\Id}{\ensuremath{\mathbb{I}}}
\newcommand{\pd}[2]{\ensuremath{\frac{\partial #1}{\partial #2}}}
\newcommand{\conj}[1]{\ensuremath{#1^{\ast}}}
\newcommand{\T}[1]{\ensuremath{#1^{\dagger}}}
\newcommand{\TT}[1]{\ensuremath{#1^{\top}}}

\newcommand{\HaarProdSet}{\ensuremath{\mc{S}_{\text{Haar}_1^{\otimes n}}}}

\newcommand{\hilbert}{\ensuremath{\mathcal{H}}}

\newcommand{\Nlayers}{\ensuremath{L}}
\newcommand{\Ngates}{\ensuremath{K}}  
\newcommand{\Nmaps}{\ensuremath{K}}
\newcommand{\As}{\ensuremath{\vb*{A}}}
\newcommand{\Gs}{\ensuremath{\vb*{G}}}
\newcommand{\Zs}{\ensuremath{\vb*{V}}}
\newcommand{\indSample}{\ensuremath{s}}
\newcommand{\indGate}{\ensuremath{k}}
\newcommand{\indMap}{\ensuremath{k}}
\newcommand{\indLayer}{\ensuremath{\ell}}
\newcommand{\indLayerList}{\ensuremath{\mc{I}_\ell}}

\newcommand{\z}{\ensuremath{\vb*{z}}}
\newcommand{\zc}{\ensuremath{\vb*{z}^\ast}}
\renewcommand{\v}{\ensuremath{\vb*{v}}}

\newcommand{\x}{\ensuremath{\vb*{x}}}
\newcommand{\y}{\ensuremath{\vb*{y}}}

\newcommand{\brickwall}{\ensuremath{G}}
\newcommand{\map}{\ensuremath{A}}
\newcommand{\A}[1]{\ensuremath{A^{[#1]}}}  
\newcommand{\AT}[1]{\ensuremath{A^{[#1]\dagger}}}  
\newcommand{\AC}[1]{\ensuremath{A^{[#1]\ast}}}  
\newcommand{\ATT}[1]{\ensuremath{A^{[#1]\top}}}  
\newcommand{\G}[1]{\ensuremath{G^{[#1]}}}  
\newcommand{\Z}[1]{\ensuremath{V^{[#1]}}}  
\newcommand{\Gl}[1]{\ensuremath{G_{(#1)}}}  
\newcommand{\Zl}[1]{\ensuremath{Z_{(#1)}}}  
\newcommand{\GT}[1]{\ensuremath{G^{[#1]\dagger}}}  
\newcommand{\ZTT}[1]{\ensuremath{V^{[#1]\top}}}  
\newcommand{\ZT}[1]{\ensuremath{V^{[#1]\dagger}}}  
\newcommand{\GlT}[1]{\ensuremath{G_{(#1)}^\dagger}}  
\newcommand{\ZlT}[1]{\ensuremath{Z_{(#1)}^\dagger}}  
\newcommand{\Gopt}{\ensuremath{\vb*{G}_\text{opt}}}  

\newcommand{\fisVar}{\ensuremath{\psi}}  
\newcommand{\bisVar}{\ensuremath{\phi}}  
\newcommand{\dbisVar}{\ensuremath{\delta\phi}}  

\newcommand{\fis}[1]{\ensuremath{\fisVar^{[#1]}}}  
\newcommand{\bis}[1]{\ensuremath{\bisVar^{[#1]}}}  
\newcommand{\bisT}[1]{\ensuremath{\bisVar^{[#1]\dagger}}}  
\newcommand{\bisTT}[1]{\ensuremath{\bisVar^{[#1]\top}}}  
\newcommand{\bisC}[1]{\ensuremath{\bisVar^{[#1]\ast}}}

\newcommand{\dbis}[1]{\ensuremath{\delta\bis{#1}}}  
\newcommand{\dfis}[1]{\ensuremath{\delta\fis{#1}}}  
\newcommand{\dbisC}[1]{\ensuremath{\dbisVar^{[#1]\ast}}}  

\newcommand{\fisVarSample}{\ensuremath{\fisVar^{(s)}}}  
\newcommand{\bisVarSample}{\ensuremath{\bisVar^{(s)}}}  

\newcommand{\bbis}[1]{\ensuremath{\Bar{\bisVar}^{[#1]}}}  
\newcommand{\bfis}[1]{\ensuremath{\Bar{\fisVar}^{[#1]}}}  

\newcommand{\Nsamples}{\ensuremath{N_s}}  

\newcommand{\Uref}{\ensuremath{U_\text{ref}}}

\newcommand{\costHSTot}{\ensuremath{C_\text{HS}}}
\newcommand{\costFTot}{\ensuremath{C_\text{F}}}
\newcommand{\costF}{\ensuremath{\mathcal{L}_\text{F}}}
\newcommand{\costHS}{\ensuremath{\mathcal{L}_\text{HS}}}
\newcommand{\costHSSample}{\ensuremath{\costHS^{(\indSample)}}}

\newcommand{\overlap}{\ensuremath{\mathcal{T}(\Gs)}}
\newcommand{\overlapVar}{\ensuremath{\mathcal{T}}}

\renewcommand{\grad}{\ensuremath{\nabla}}   
\newcommand{\gradi}[1]{\ensuremath{\grad_{#1}}}
\newcommand{\hvp}[2]{\ensuremath{H\!\left( #1\right)[ #2 ]}}
\newcommand{\dgrad}[2]{\ensuremath{D\left( #1\right)[ #2 ]}}



%% file: 0_abstract.tex
\begin{abstract}
Optimizing tensor networks with standard first-order methods often leads to slow convergence and getting trapped in local minima.
Although second-order optimization offers enhanced robustness, explicitly constructing the full Hessian matrix is computationally prohibitive for large-scale systems. 
In this work, we bypass this bottleneck by introducing an analytical Hessian-vector product kernel designed for arbitrary compositions of linear maps.
This two-pass algorithm leverages a recursive tangent-state propagation with a bounded virtual bond dimension to guarantee scalability. 
We demonstrate its use within a Riemannian trust-region method for quantum circuit compression.
For time-evolution circuits with up to 17 layers, our method reproduces previous results in substantially fewer iterations, reducing wall-time by up to an order of magnitude.
For deeper quantum circuits, incorporating second-order derivative information yields a better solution, reducing the final error to less than half that of the first-order result.
\end{abstract}

%% file: 1_introduction.tex
\section{Introduction \label{sec:introduction}}

\Ac{TN} methods provide a powerful variational framework for high‑dimensional Hilbert spaces by decomposing large tensors into networks of computationally manageable components.
Originally developed in condensed matter physics to study strongly correlated materials~\cite{white1992density,schollwock2011density,orus2014practical}, their utility now extends to quantum chemistry~\cite{chan2011density,chan2016matrix}, machine learning~\cite{stoudenmire2016supervised,reyes2021multi}, and quantum circuit simulation~\cite{markov2008simulating,seitz2023simulating}. 
By efficiently encoding entanglement and correlation structures, \acp{TN} mitigate the \emph{curse of dimensionality} inherent to many-body problems.

The central challenge in these applications is optimizing \ac{TN} parameters to accurately represent a target system, typically by minimizing a cost function defined over network states. 
Although a variety of specialized algorithms exist~\cite{white1992density, haegeman2011time, haegeman2016unifying, zauner2018variational}, gradient-descent methods remain a popular choice for many applications~\cite{stoudenmire2016supervised, le2025riemannian, zhang2024scalable}. 
However, since these methods rely solely on local gradient information, they are prone to local minima and converge slowly in high-dimensional optimization landscapes. 
These limitations motivate the exploration of second-order optimization techniques~\cite{absil2008optimization, haegeman2013post, vanderstraeten2016gradient, kotil2024riemannian, putterer2025high, ben2025regularized}, which incorporate curvature information to accelerate convergence and improve robustness.

\begin{figure}
    \centering
    \includegraphics[width=0.9\linewidth]{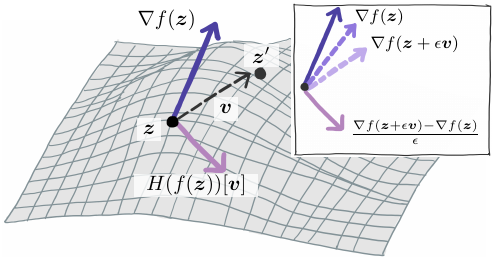}
    \caption{Visualization of the gradient and its directional derivative, i.e., the Hessian-vector product $\hvp{f(\z)}{\v}$. The gradient $\grad f$ represents the local slope; the difference between the gradient at point $\z$ and a perturbed point $\z + \epsilon \v$ defines the Hessian-vector product as $\epsilon \to 0$.}
    \label{fig:derivatives}
\end{figure}

A key ingredient of second-order methods is access to curvature information via the Hessian matrix $\vb*{H}$. 
Since forming $\vb*{H}$ explicitly is typically infeasible due to its quadratic scaling with the number of parameters, practical algorithms instead rely on \acp{HVP}~\cite{nocedal2006numerical,pearlmutter1994fast,griewank2008evaluating}. These compute the action of the Hessian on a vector $\vb*{v}$, denoted as $\hvp{f(\z)}{\v}$, without constructing the full matrix (\cref{fig:derivatives}).

In principle, \acp{HVP} can be evaluated efficiently using \ac{AD} techniques~\cite{pearlmutter1994fast,griewank2008evaluating}, achieving computational costs comparable to a single gradient evaluation. 
Modern \ac{AD} frameworks have enabled differentiable \ac{TN} programming~\cite{liao2019differentiable,francuz2025stable}, offering considerable flexibility by treating \ac{TN} contractions as generic computational graphs. 
However, this generality makes \ac{AD} agnostic to the \ac{TN} structure.
Since it differentiates the truncated algorithm itself~\cite{liao2019differentiable}, truncation steps such as \acp{SVD} enter the computational graph and require carefully stabilized derivative rules~\cite{liao2019differentiable,francuz2025stable}.
Additionally, these must be differentiated once more for an \ac{HVP}.
Here, we propose a method that avoids differentiating the truncation by propagating a so-called tangent state explicitly as an \ac{MPS} and truncating it by means of a projection onto the MPS tangent space.

Within the \ac{TN} community, second-order information has mostly been explored in specialized contexts.
For uniform \ac{MPS} in the thermodynamic limit, the tangent-space formalism~\cite{haegeman2011time,haegeman2013post} provides a global framework, which includes an explicit construction of the Hessian of the energy functional to characterize elementary excitations. The tangent states used in this work build on this concept and correspond to the generalized tangent vectors of Ref.~\cite{haegeman2013post}, evaluated along a single direction for a finite composition of linear maps.
Regularized analytical Hessians have also been formulated for training Born machines~\cite{ben2025regularized}; however, they are restricted to single-tensor updates under a sweeping scheme.
Both approaches are thus tailored to a specific objective and rely either on translation invariance or on site-local updates, rather than providing global second-order information for general \ac{TN} contractions and any cost function that depends on the network through an overlap.

In this work, we consider arbitrary compositions of linear maps -- a structure fundamental to \acp{TN}, and 
derive an analytical \ac{HVP} kernel based on the principles of \ac{AD}.
By formulating it directly in terms of \ac{TN} contractions, the bond dimension of all intermediate states can be controlled explicitly, while the tangent states remain representable using at most twice that bond dimension.
Our approach avoids differentiating through truncation steps, making the kernel applicable to large approximate \ac{TN} contractions.
It remains agnostic to any cost function that depends on the network via the overlap, and it avoids explicit construction of the Hessian.

We demonstrate the kernel for quantum circuit compression, which connects two of our previous works.
Ref.~\cite{kotil2024riemannian} established the Riemannian trust-region framework with truncated \ac{CG}, including the Riemannian Hessian on the unitary manifold.
Because the Hessian was explicitly constructed as a dense matrix, the optimization was limited to small systems, with larger systems reached only by exploiting translation invariance.
Ref.~\cite{le2025riemannian} developed a scalable first-order scheme based on \acp{TN}.
The present work provides the missing component: a scalable, matrix-free \ac{HVP} that enables the trust-region framework of Ref.~\cite{kotil2024riemannian} to be applied to system sizes accessible to Ref.~\cite{le2025riemannian}.

%% file: 2_general-method.tex
\section{General method: \acp{HVP} for compositions of linear maps \label{sec:general-method}}

We consider systems whose underlying computational graph is a composition of linear maps. 
In this section, we first formalize the general concept of the \ac{HVP} and subsequently establish a systematic framework for its efficient evaluation within such composite functions.

\subsection{Complex derivatives}

To characterize the optimization landscape for functions of complex variables, we must establish a consistent derivative framework over complex vector spaces. 
We employ the Wirtinger formalism, which treats a complex vector $\z=\x+i\y\in\C^d$ and its conjugate $\zc$ as formally independent variables. 
This approach enables elegant differentiation of non-holomorphic functions without decomposing them into real and imaginary components.
A concise summary of this formalism is provided in \cref{ap:wirtinger}; for a comprehensive treatment, we refer the reader to Ref.~\cite{koor2023short}.
The Wirtinger derivatives are defined in terms of the standard real partial derivatives as:
\begin{align}
\label{eq:wirtinger-derivatives}
    \pd{f}{\z} = \frac{1}{2} \pa{\pd{f}{\x} - i\pd{f}{\y}}, \quad
    \pd{f}{\zc} = \frac{1}{2} \pa{\pd{f}{\x} + i\pd{f}{\y}}.
\end{align}

The first-order variation of a function $f$ along a vector $\v\in\C^d$ is captured by the directional derivative given by (with $\epsilon\in\R$):
\begin{equation}
    \label{eq:directional-derivative-definition}
    \dgrad{f(\z)}{\v} = \lim_{\epsilon \to 0} \frac{f(\z + \epsilon \v) - f(\z)}{\epsilon}.
\end{equation}
Note that this definition extends naturally to vector-valued functions $\vb*{f}:\C^d\rightarrow\C^m$.

While the directional derivative describes the rate of change in a specific direction, optimization requires a gradient vector to identify the path of steepest ascent. 
The definition of this vector depends on the nature of $f$:
\begin{enumerate}[(a)]
\item \textit{Real-valued functions:} For $f_r:\C^d\rightarrow\R$, the complex gradient is defined as:
\begin{equation}
    \label{eq:grad-real}
    \grad f_r(\z) = 2 \cdot \conj{\pa{\pd{f_r}{\z}}} \in \C^d.
\end{equation}
\item \textit{Holomorphic functions:} For holomorphic $f_h:\C^d\rightarrow\C$, the conjugate derivative vanishes ($\partial f_h / \partial \zc = 0$). In this case, the first-order behavior is fully described by the complex derivative vector:
\begin{equation}
    \label{eq:grad-holom}
    \grad f_h(\z) = \pd{f_h}{\z} \in \C^d.
\end{equation}
\end{enumerate}

Building upon these first-order primitives, we can construct the \ac{HVP}, which extracts second-order information in a specific direction $\v$ without the memory and computational costs of explicitly constructing the full Hessian matrix. 
Provided $f$ is either real-valued or holomorphic, the symmetry of mixed partial derivatives allows us to evaluate the \ac{HVP} by nesting first-order operations in two equivalent modes~\cite{griewank2008evaluating, pearlmutter1994fast}:
\begin{enumerate}[(i)]
\item \textit{Reverse-over-reverse mode} (gradient of the directional derivative):
\begin{equation}
\label{eq:RoR}
\hvp{f(\z)}{\v} = \grad \br{\dgrad{f(\z)}{\v}} \in \C^d
\end{equation}
\item \textit{Forward-over-reverse mode} (directional derivative of the gradient):
\begin{equation}
\label{eq:FoR}
\hvp{f(\z)}{\v} = \br{\dgrad{\grad_j f(\z)}{\v}}_{j=1 \dots d} \in \C^d
\end{equation}
\end{enumerate}

\subsection{The holomorphic overlap}

We now focus on systems where the computational graph is structured as a sequential composition of linear operations. 
Let $\As = (\A{1}, \ldots, \A{\Ngates})$ be a sequence of complex-valued linear maps on the Hilbert space $\hilbert$, i.e., $\A{\indMap} \in \mc{L}(\hilbert)$. 
The composition gives the total evolution:
\begin{equation}
    \map = \prod_{\indMap = \Nmaps}^1 \A{\indMap} = \A{\Nmaps} \cdots \A{2} \cdot \A{1}.
\end{equation}
The objective is to analyze the scalar overlap $\overlapVar(\As)$ between the evolved state $\map\fisVar$ and a reference state $\bisVar\in\hilbert$:
\begin{equation}
    \overlapVar(\As) = \T{\bisVar} \map \fisVar = \T{\bisVar} \pa{\A{\Nmaps} \cdots \A{1}} \fisVar,
\end{equation}
where $\T{(\cdot)}$ denotes the adjoint.

We treat the matrix entries $A_{ij}^{[k]}\in\C$ as independent complex parameters. In terms of these, $\overlapVar(\As)$ is a polynomial in the entries that contains none of their complex conjugates. It is therefore a holomorphic function of the map parameters, and all conjugate Wirtinger derivatives vanish:
\begin{equation}
    \pd{\overlapVar}{\AC{\indMap}} = 0 \quad \text{for all } \indMap = 1, \dots, \Nmaps.
\end{equation}
This holomorphicity simplifies the derivation of higher-order derivatives, since the first-order behavior is fully captured by the complex derivative vector defined in \cref{eq:grad-holom}.

\subsection{Forward and backward passes \label{sec:general-computational-passes}}

Evaluating $\overlapVar(\As)$ and its derivatives involves treating the composition as a bidirectional computational graph. 
We define a sequence of \emph{intermediate states} by propagating boundary information through the network from both ends. 
In the \emph{forward pass}, the initial state $\fisVar$ is evolved through the sequence of maps; in the \emph{backward pass}, the reference state $\bisVar$ is propagated in reverse using the adjoint operators:
\begin{subequations}
\begin{align}
    \fis{0} &= \fisVar, &\quad \fis{\indMap} &= \A{\indMap} \fis{\indMap-1}, \label{eq:fis} \\
    \bis{0} &= \bisVar, &\quad \bis{\indMap} &= \AT{\Nmaps-\indMap+1} \bis{\indMap-1}, \label{eq:bis}
\end{align}
\end{subequations}
for $\indMap=1,\ldots,\Nmaps$. 
Here, $\fis{\indMap}$ represents the forward state after the application of the $\indMap$-th linear map, while $\bis{\indMap}$ represents the backward ``co-state'' evolved from the reference side.

At any intermediate depth $\indMap$, the total overlap can be recovered by contracting the corresponding forward and backward states:
\begin{align}
\label{eq:overlap-general}
    \overlapVar(\As) = \bisT{\Nmaps-\indMap} \fis{\indMap}.
\end{align}
This expression shows that the overlap is independent of the contraction point $k$, so it can be evaluated in a single forward or backward sweep.
Importantly, these intermediate states serve as the building blocks for the first-order gradients and the second-order \acp{HVP} derived in the following subsections, as they effectively ``cache'' the environment of each map $\A{\indMap}$.

\subsection{The complex derivatives \label{sec:general-complex-gradient}}

To determine the Wirtinger derivatives, we isolate the contribution of a specific local map $\A{\indMap}$ to the total overlap:
\begin{align}
\label{eq:general-map-isolation}
    \overlapVar(\As) = \bisT{\Nmaps-\indMap} \A{\indMap} \fis{\indMap-1}.
\end{align}
As established, the holomorphic nature of the overlap ensures that the conjugate Wirtinger derivatives vanish:
\begin{equation}
    \pd{\overlapVar(\As)}{\AT{\indMap}} = 0.
\end{equation}
The derivative with respect to the map parameters $\A{\indMap}$ is given by the outer product of the backward and forward intermediate states:
\begin{equation}
\label{eq:general-doverlap}
    \pd{\overlapVar(\As)}{\A{\indMap}} = \bisC{\Nmaps - \indMap} \otimes \fis{\indMap-1}.
\end{equation}

Computationally, evaluating these complex derivatives involves a bidirectional sweep: one first performs a forward pass to cache the states $\fis{\indMap}$, followed by a backward pass to iteratively compute the states $\bis{\indMap}$ and perform the outer product at each depth $\indMap$.
Alternatively, the order can be reversed by caching the backward states first.

\subsection{Recursive tangent-state propagation}

To evaluate the sensitivity of the final overlap, we consider a collective variation of the maps along the direction of tangent vectors $\Zs = (\Z{1},\ldots,\Z{\Nmaps})$.
We define the varied maps as:
\begin{equation}
    \A{\indMap}(\epsilon) = \A{\indMap} + \epsilon \Z{\indMap}, \qquad \epsilon\in\R,
\end{equation}
for $\indMap = 1\ldots\Nmaps$.
This shift in the constituent maps induces a corresponding variation in the intermediate forward and backward states, which is propagated sequentially:
\begin{align*}
    \fis{0}(\epsilon) &= \fisVar,  \qquad &\fis{\indMap}(\epsilon) &= \A{\indMap}(\epsilon)\,\fis{\indMap-1}(\epsilon), \\
    \bis{0}(\epsilon) &= \bisVar,  \qquad &\bis{\Nmaps-\indMap+1}(\epsilon) &= \AT{\indMap}(\epsilon)\,\bis{\Nmaps-\indMap}(\epsilon).
\end{align*}
The directional derivative of the forward intermediate state, $\dfis{\indMap} \coloneq \dgrad{\fis{\indMap}}{\Zs}$, represents the exact accumulated variation up to step $\indMap$. 
Following the definition of the directional derivative from \cref{eq:directional-derivative-definition}, we obtain:
\begin{equation}
\begin{split}
    \dfis{\indMap} &= \lim_{\epsilon\to0} \frac{\fis{\indMap}(\epsilon) - \fis{\indMap}}{\epsilon} \\
    &= \lim_{\epsilon\to0} \pa{\A{\indMap} \frac{\fis{\indMap-1}(\epsilon) - \fis{\indMap-1}}{\epsilon} + \Z{\indMap}\fis{\indMap-1}(\epsilon)} \\
    &= \A{\indMap} \dfis{\indMap-1} + \Z{\indMap} \fis{\indMap-1}.
\end{split}
\end{equation}
An analogous derivation yields the directional derivatives for the backward intermediate states, $\dbis{\indMap} \coloneq \dgrad{\bis{\indMap}}{\Zs}$. 
Since the boundary states $\fis{0}$ and $\bis{0}$ are fixed and independent of the local maps, the initial conditions for this propagation are $\dfis{0}=0$ and $\dbis{0}=0$.
In summary, the recursive update rules for the \emph{recursive tangent-state propagation} are given as follows:
\begin{subequations}
    \label{eq:general-tangent-states-recursion}
    \begin{align}
        \dfis{0} &= 0, 
         &\dfis{\indMap} &= \A{\indMap}\dfis{\indMap-1} + \Z{\indMap} \fis{\indMap-1}, \label{eq:dfis}\\
         \dbis{0} &= 0, 
        &\dbis{\indMap} &= \AT{\Nmaps-\indMap+1}\dbis{\indMap-1} + \ZT{\Nmaps-\indMap+1} \bis{\indMap-1}. \label{eq:dbis}
    \end{align}
\end{subequations}

\subsection{The Hessian-vector product}

Following the forward-over-reverse mode given in \cref{eq:FoR}, the \ac{HVP} for the holomorphic overlap can be computed as:
\begin{align*}
    \hvp{\overlapVar(\As)}{\Zs} = \dgrad{\grad\overlapVar(\As)}{\Zs},
\end{align*}
with the directional derivative given by \cref{eq:directional-derivative-definition}.
Since the overlap is holomorphic, $\grad\overlapVar(\As)$ is simply the complex derivative defined in \cref{eq:general-doverlap}.
We note that in the latter equation, $\bisC{\Nmaps-\indMap}$ is a function of $\As^\top$ instead of $\As^\dagger$ and hence the product rule (see \cref{ap:eq:dgrad-product-vec}) can be applied.
Using the \tangent states as presented in \cref{eq:general-tangent-states-recursion}, we obtain:
\begin{equation}
\label{eq:overlap-hvp}
\begin{aligned}
    \hvp{\overlapVar(\As)}{\Zs} &= \dbisC{\Nmaps-\indMap} \otimes \fis{\indMap-1} \\
    &\ \, + \bisC{\Nmaps-\indMap} \otimes \dfis{\indMap-1}.
\end{aligned}
\end{equation}
This formulation reveals an algebraic symmetry: the local curvature at $\A{\indMap}$ is a sum of two distinct outer products. 
The first term, $\dbisC{\Nmaps-\indMap} \otimes \fis{\indMap-1}$, aggregates the variation of all maps occurring \emph{after} $\A{\indMap}$ in the computational graph. 
The second term, $\bisC{\Nmaps-\indMap} \otimes \dfis{\indMap-1}$, captures the accumulated variation of all maps \emph{preceding} $\A{\indMap}$.

As is well known in the \ac{AD} community, \ac{HVP} can be obtained in various ways, e.g., by reverse-over-reverse or forward-over-reverse mode, see \cref{eq:FoR,eq:RoR}.
While all these methods compute the same quantity, they differ in the order of operations and in memory when applied  as black-box \ac{AD}.
Here, however, we derive an explicit kernel for the \ac{HVP} based on these principles instead of applying them as a black box.
Since the adjoint of the adjoint model is the tangent-linear model~\cite{christianson1992automatic,griewank2008evaluating}, one expects both derivations to lead to the same kernel, which we indeed verify explicitly in \cref{ap:RoR-derivation}. 

\begin{figure}
    \centering
    \includegraphics[width=0.9\linewidth]{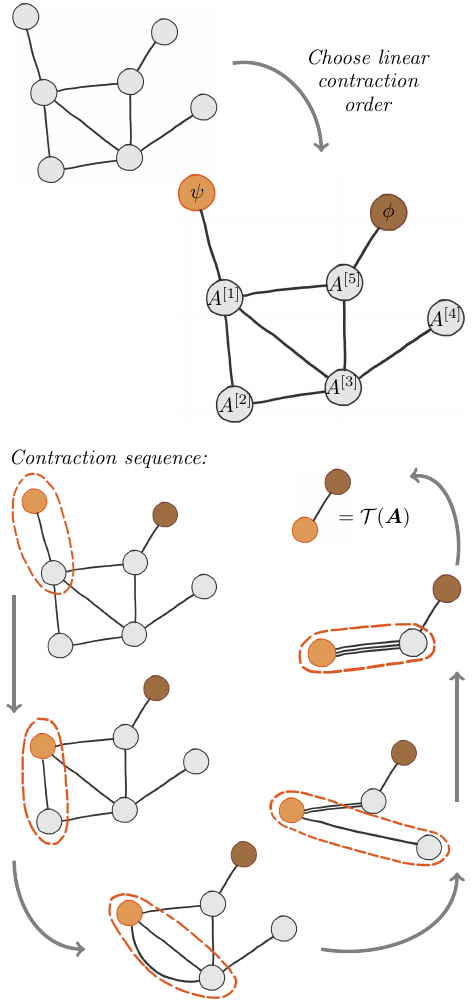}
    \caption{\Ac{TN} representation as a composition of linear maps. By choosing a linear contraction order, a general \ac{TN} can be interpreted as a sequential composition of local tensors $\As$ and boundary states $\fisVar,\,\bisVar$.}
    \label{fig:general-TN}
\end{figure}

\subsection{Algorithm to evaluate derivatives of the overlap\label{sec:algorithm-overlap}}

We consolidate the preceding derivations into a unified algorithm that evaluates the overlap $\overlapVar$, its gradient $\grad\overlapVar$, and the directional curvature $\hvp{\overlapVar}{\Zs}$ through a single pair of forward and backward sweeps. 
The execution order is symmetric: one may perform the forward pass first, followed by the backward pass, or vice versa.

In the implementation detailed here, the initial forward pass caches the intermediate states $\fis{\indMap}$ and their corresponding tangent variations $\dfis{\indMap}$. 
During the subsequent backward pass, the remaining backward states $\bis{\indMap}$ and their tangent counterparts $\dbis{\indMap}$ are computed on the fly. 
This concurrent execution allows the simultaneous accumulation of the gradient $\grad\overlapVar$ and the \ac{HVP} kernel.
The complete procedure, structured for an initial forward pass followed by a backward sweep, is presented in \cref{alg:hvp-kernel-core}.
\begin{figure*}
\begin{minipage}{\linewidth}
\begin{algorithm}[H]
\caption{Analytical \ac{HVP} kernel for sequential linear map compositions}\label{alg:hvp-kernel-core}
\begin{algorithmic}[1]
\Require $\fis{0}, \bis{0}, \As, \Zs$
\Procedure{HVP\_Kernel}{$\fis{0}, \bis{0}, \As, \Zs$}
    \Procedure{Forward pass}{$\fis{0}, \As, \Zs$}
        \For{$\indMap = 1, \dots, \Nmaps$}
        \State $\dfis{\indMap} \gets \A{\indMap} \dfis{\indMap-1} + \Z{\indMap} \fis{\indMap -1}$ \Comment{Compute and cache the forward \tangent states, \cref{eq:dfis}}
        \State $\fis{\indMap} \gets \A{\indMap}\fis{\indMap-1}$ \Comment{Compute and cache the forward intermediate state, \cref{eq:fis}}
        \EndFor
    \EndProcedure
    \Procedure{Backward pass}{$\bis{0}, \{ \fis{\indMap} \}_{\indMap=1\dots \Nmaps}, \{\dfis{\indMap}\}_{\indMap=1\dots \Nmaps}, \As, \Zs$}
        \State $\bisVar \gets \bis{0}$
        \State $\dbisVar \gets 0$
        \For{$\indMap = \Nmaps, \dots, 1$}
        \State $\dbisVar \gets \AT{\indMap} \dbisVar + \ZT{^\indMap} \bisVar$ \Comment{Update the backward \tangent state, \cref{eq:dbis}}
        \State $\grad_{\indMap} \overlapVar(\As) \gets \bisC{\Nmaps-\indMap} \otimes \fis{\indMap-1}$  \Comment{Build up $\grad\overlapVar$, \cref{eq:general-doverlap}}
        \State $\br{\hvp{\overlapVar(\As)}{\Zs}}_\indMap \gets \dbisC{\Nmaps-\indMap} \otimes \fis{\indMap-1} + \bisC{\Nmaps-\indMap} \otimes \dfis{\indMap-1}$  \Comment{Accumulate \ac{HVP} of $\overlapVar$, \cref{eq:overlap-hvp}}
        \State $\bisVar \gets \AT{\indMap} \bisVar$ \Comment{Update the backward intermediate state, \cref{eq:bis}}
        \EndFor
        \State $\overlapVar(\As) \gets \T{\bisVar} \fis{0}$ \Comment{Compute $\overlapVar$ at boundary, \cref{eq:overlap-general}}
    \EndProcedure
\EndProcedure
\Ensure $\overlapVar, \grad\overlapVar, \hvp{\overlapVar}{\Zs}$
\end{algorithmic}
\end{algorithm}
\end{minipage}
\end{figure*}

%% file: 3_quantum-setting.tex
\section{Application to tensor networks and quantum circuits \label{sec:quantum-setting}}

Computationally, the recursive state propagation and overlap evaluations detailed in \cref{sec:general-method} map directly onto specific \ac{TN} contraction sequences. 
By defining a linear contraction order, any \ac{TN} composed of local tensors $\As$ and fully contractible to a scalar can be represented as a composition of linear maps.
This structural mapping, visualized in \cref{fig:general-TN}, allows the general \ac{HVP} formalism to be applied seamlessly to \ac{TN} architectures. 
Consequently, the network's second-order information can be extracted via a sequence of local maps, preserving the scalability inherent to the recursive approach.

To demonstrate the practical utility of this framework, we now specialize the formalism to the optimization of quantum circuits. 
This setting provides a natural domain for the analytical \ac{HVP} kernel, in which the constituent operations and their derivatives are expressed using intuitive tensor diagrams. 
In the following, we adopt a \ac{MPS} representation for the quantum states; however, we emphasize that the kernel's underlying logic remains applicable to any \ac{TN} geometry that admits a linear decomposition into sequential maps.

In this context, the local maps correspond to a sequence of quantum gates $\Gs = (\G{1}, \dots, \G{\Ngates})$. 
Their composite action defines the quantum circuit $\brickwall \in \unitary(d)$ acting on a $d$-dimensional Hilbert space. 
Since the quantum gates $\Gs$ are constrained to the unitary manifold, the derivatives need to lie in the corresponding tangent spaces. 
To maintain simplicity, the following section focuses on Euclidean derivatives, noting that these can be transformed via Riemannian projections as described in \cref{ap:projection}.

\subsection{Problem setting}

When optimizing quantum circuits, it is crucial to quantify the distance between quantum evolutions given by a target unitary $\Uref \in \unitary(d)$ and an ansatz $\brickwall \in \unitary(d)$. 
While our methodology applies to arbitrary layouts, we demonstrate the utility using a brickwall circuit $\brickwall$ parameterized by a sequence of two-qubit gates $\Gs \in \unitary(4)^{\times \Ngates}$ to provide an explicit visualization.

To measure the distance between $\brickwall$ and $\Uref$, we employ the Hilbert-Schmidt test \cite{khatri2019quantum} and the Frobenius norm:
\begin{align}
    \label{eq:HST}
    \costHSTot(\Uref, \brickwall) &= \mathbb{E}_{\fisVar \sim \text{Haar}} \br{ 1 - \abs{\bra{\fisVar} \Uref^{\dagger} \brickwall \ket{\fisVar}}^2 }, \\
    \label{eq:Frobenius}
    \costFTot(\Uref, \brickwall) &= \mathbb{E}_{\fisVar \sim \text{Haar}} \br{ 1 - \Re\br{\bra{\fisVar} \Uref^{\dagger} \brickwall \ket{\fisVar}}}.
\end{align}
In practice, these Haar expectations are approximated using a finite set of sample states $\{ \fisVarSample \}_{s=1}^{\Nsamples}$ \cite{caro2023out, zhang2024scalable}. 
For a single sample $\fisVar$, we define the reference state $\bisVar = \Uref\fisVar$ and identify the core overlap as $\overlap = \T{\bisVar} \brickwall \fisVar$. 
The sample-wise variational contributions to the total cost are then captured by:
\begin{equation}
\label{eq:cost-summands}
    \costHS = \abs{\overlap}^2, \quad \costF = \Re\br{\overlap}.
\end{equation}
Now, $\costHS$ and $\costF$ are both real-valued functions but not holomorphic anymore, as they additionally depend on the complex conjugate parameters $\conj{\Gs}$. 

Since we specifically represent the states $\fisVar$ and $\bisVar$ as \acp{MPS}, the overlap $\overlap$ is naturally visualized as a \ac{TN} contraction:
\begin{equation}
\label{eq:overlap-plain}
\mathcal{T}(\Gs) = 
\begin{minipage}[h]{4cm}
\includegraphics[width=4cm]{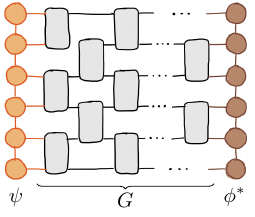}
\end{minipage}.
\end{equation}
Note that the \ac{TN} diagrams use a ``left-to-right''-ordering convention, different from the ``right-to-left''-ordering of matrix-vector multiplications. 

To optimize these circuits, we require first- and second-order derivatives of $\costHS$. 
Following Ref.~\cite{le2025riemannian}, the complex gradient $\grad\costHS$ can be related to the gradient of $\costF$ as follows:
\begin{equation}
    \label{eq:grad-costHS}
    \grad\costHS(\Gs) = -2 \overlap \cdot \grad\costF(\Gs).
\end{equation}
Beyond the gradient, the \ac{HVP} extracts the second-order curvature along an arbitrary variational direction $\Zs$ in the tangent space. The derivation below holds for any $\Zs$. Its choice is determined by the optimization algorithm that calls the \ac{HVP}; in the trust-region method of \cref{sec:riemannian-opt}, the directions are generated by the truncated \ac{CG} method~\cite[Algorithm 11]{absil2008optimization}.
Applying the product rule (\cref{ap:eq:dgrad-product}) to \cref{eq:grad-costHS} allows us to compute the \ac{HVP} as the directional derivative of the complex gradient:
\begin{equation}
    \label{eq:hvp-costHS} 
    \hvp{\costHS(\Gs)}{\Zs} = 2 \br{ \Omega \grad\costF(\Gs) + \overlap \hvp{\costF(\Gs)}{\Zs}},
\end{equation}
where $\Omega \coloneq  \dgrad{\overlap}{\Zs}$ is the directional derivative of the overlap. 
This decomposition shows that the second-order geometry of the quadratic Hilbert-Schmidt test is fully determined by the first- and second-order derivatives of the linear Frobenius norm.
In the following subsections, we leverage this notation to derive efficient kernels for these derivatives, while maintaining a consistent \ac{TN} interpretation of the underlying contractions.

\begin{figure}[H]
    \centering
    \includegraphics[width=\linewidth]{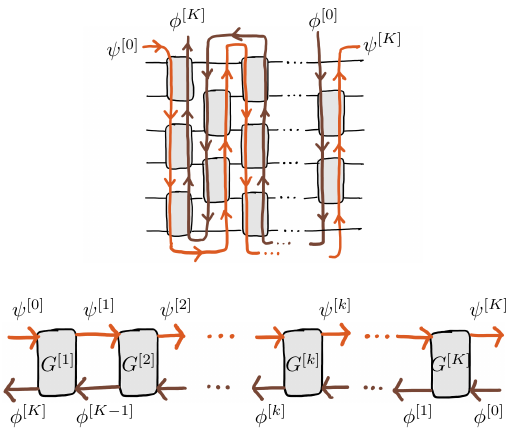}
    \caption{Computational passes for a brickwall circuit. Forward states $\fis{\indGate}$ are generated by sequentially applying gates $\G{\indGate}$, while backward states $\bis{\indGate}$ are computed by applying the adjoint gates $\GT{\indGate}$ in reverse order, starting from the reference state.}
    \label{fig:computational-passes}
\end{figure}

\subsection{Computational passes}

The bidirectional information flow introduced in \cref{sec:general-computational-passes} is here tailored to the brickwall circuit architecture in \cref{fig:computational-passes}~\cite{putterer2025high}. 
The resulting forward and backward intermediate states, defined as in \cref{eq:fis,eq:bis}, form the computational building blocks for our analytical derivatives. 
These states effectively encapsulate the system's ``memory'' of the evolution at each circuit depth, providing the local environments necessary for exact gradient and curvature evaluations.

If the initial gates are chosen such that $\brickwall$ already approximates $\Uref$, e.g., by choosing an initial Trotter circuit, the bond dimension of the intermediate forward and backward state is capped by the maximum bond dimension $\chi$ of the reference state $\bisVar$ since $\brickwall$ effectively disentangles $\Uref$~\cite{gibbs2024deep}.

\subsection{Evaluating the overlap in a single pass}

Isolating the action of a specific gate $\G{\indGate}$ as in \cref{eq:general-map-isolation} allows us to express the global overlap $\overlap$ as a local contraction between the gate and its environment. 
To facilitate this, we distinguish the physical indices $(i,j)$ coupled by $\G{\indGate}$ from the unaffected indices $\alpha$ that bypass the local gate:
\begin{align}
\label{eq:overlap-2}
    \overlap &= \sum_{ij\alpha} \bisC{\Ngates-\indGate}_{i\alpha} \G{\indGate}_{ij} \fis{\indGate-1}_{j\alpha} \\
    &= 
    \begin{minipage}[h]{7.5cm}
    \vspace{0pt}
    \includegraphics[width=7.5cm]{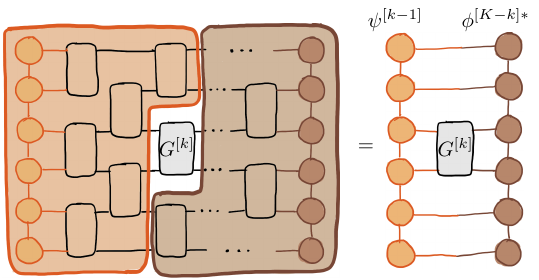}
    \end{minipage}. \nonumber
\end{align}
Consistent with the bidirectional flow in \cref{sec:general-computational-passes}, the total overlap $\overlap$ is invariant to the choice of $\indGate$ and can be retrieved at any circuit depth. 
Consequently, evaluating $\overlap$ requires only a single forward or backward pass, as it is equivalent to the boundary contractions $\bisT{\Ngates} \fis{0}$ or $\bisT{0} \fis{\Ngates}$.

\subsection{The complex gradient and gate environment \label{sec:complex-gradient}}

Following Ref.~\cite{le2025riemannian}, the complex gradient of $\costF$ is given by:
\begin{align}
    \label{eq:grad-costF}
    \grad\costF(\Gs) = -\grad\Re[\overlap] = - \conj{\br{\pd{\overlap}{\G{\indGate}}}}_{\indGate = 1\dots\Ngates}.
\end{align}
Within the \ac{TN} formalism, the Wirtinger derivative $\pd{\overlap}{\G{\indGate}}$ corresponds exactly to the environment of gate $\G{\indGate}$. 
This environment is obtained by ``cutting out'' the local gate in \cref{eq:overlap-2} and contracting the remaining network:
\begin{align}
    \label{eq:environment}
    \pd{\overlap}{\G{\indGate}_{ij}} = \sum_{\alpha} \bisC{\Ngates-\indGate}_{i\alpha} \fis{\indGate-1}_{j\alpha} = 
    \begin{minipage}[h]{2.5cm}
    \vspace{0pt}
    \includegraphics[width=2.5cm]{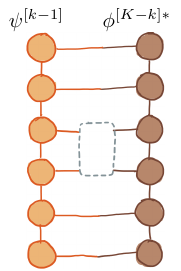}
    \end{minipage}.
\end{align}
Algebraically, this represents a partial outer product between the forward state $\fis{\indGate-1}$ and the complex-conjugated backward state $\bisC{\Ngates-\indGate}$, as established in \cref{sec:general-complex-gradient}. 
Diagrammatically, this can be interpreted as the transpose of a partial inner product over the indices $\alpha$, which is a more intuitive interpretation for the \ac{TN} diagram.

As presented in \cref{eq:grad-costHS}, the complex gradient of $\costHS$ can be constructed from $\overlap$ and $\grad \costF$.
Since $\overlap$ is available from the state propagation and $\grad\costF$ requires the contraction of the intermediate states obtained from the forward and backward passes, the full gradient $\grad\costHS$ is efficiently computed within a two-pass budget.

\subsection{Directional derivative of the overlap}

The directional derivative $\Omega$ characterizes the first-order response of the global overlap to a collective variation along $\Zs$. 
As derived in \cref{ap:Omega}, $\Omega$ is the linear superposition of local Hilbert-Schmidt inner products, $\langle A,B\rangle = \text{Tr}(A^{\dagger}B)$:
\begin{equation}
\begin{aligned}
\label{eq:Omega}
    \Omega \equiv \dgrad{\overlap}{\Zs} &= -\sum_{\indGate=1}^{\Ngates} \langle \gradi{\indGate}\costF(\Gs), \Z{\indGate} \rangle \\
    &= \sum_{\indGate=1}^{\Ngates} \conj{\br{
    \begin{minipage}[h]{2.5cm}
    \includegraphics[width=2.5cm]{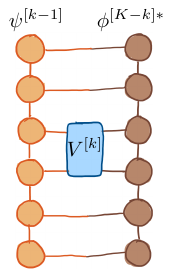}
    \end{minipage}
    }}.
\end{aligned}
\end{equation}
Diagrammatically, each term in the sum represents a circuit contraction where one original gate $\G{\indGate}$ is replaced by its variation counterpart $\Z{\indGate}$ while all other gates remain fixed for every gate. 
However, this \ac{TN} contraction does not need to be performed explicitly.
Once the gradient $\grad\costF$ is available, evaluating $\Omega$ is computationally trivial, as it only requires local $4 \times 4$ matrix contractions between the precomputed environments and the perturbation tensors.

\subsection{\Ac{HVP} via recursive \tangent-state propagation}

The \ac{HVP} of the Frobenius cost is the directional derivative of its complex gradient:
\begin{equation}
\label{eq:hvp-costF}
\begin{aligned}
    \hvp{\costF(\Gs)}{\Zs} &= \dgrad{\grad(\costF)}{\Zs} \\
    &= -\conj{\br{\dgrad{\pd{\overlap}{\G{\indGate}}}{\Zs}}}_{\indGate = 1 \dots \Ngates}.\\
\end{aligned}
\end{equation}
The essential operation is differentiating the environment of gate $\G{\indGate}$ given by \cref{eq:environment} with respect to the collective variation $\Zs$. 
Mathematically, this involves a summation over all remaining gates $\G{\indGate'}$, for $\indGate' \neq \indGate$, where in each summand a single gate $\G{\indGate'}$ is replaced by its variation counterpart $\Z{\indGate'}$. 
Following the recursive tangent-state propagation derived in \cref{eq:overlap-hvp}, we evaluate this sum by contracting intermediate states with their \tangent counterparts. 
This yields the following local expression for the directional derivative of the environment:
\begin{equation}
\begin{aligned}
    &\br{\dgrad{\pd{\overlap}{\G{\indGate}}}{\Zs}}_{ij} \\
    &\;\;\;\;\; = \underbrace{\sum_\alpha \dbisC{\Ngates-\indGate}_{i\alpha} \, \fis{\indGate-1}_{j\alpha}}_{\text{future variations}} + \underbrace{\sum_\alpha \bisC{\Ngates-\indGate}_{i\alpha} \, \dfis{\indGate-1}_{j\alpha}}_{\text{past variations}} \\
    &\;\;\;\;\;= \begin{gathered} \includegraphics[width=5.5cm]{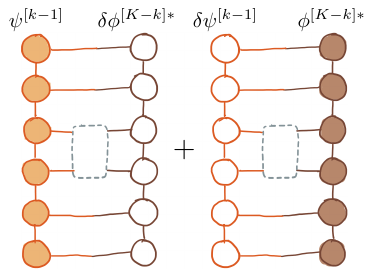} \end{gathered}.
\end{aligned}
\end{equation}
In this context, the \tangent states $\dfis{\indGate}$ and $\dbis{\indGate}$ recursively accumulate the first-order responses from the ``past'' and ``future'' variations relative to gate $\indGate$:
\begin{subequations}
\label{eq:Dpsi}
\begin{align}
    \dfis{\indGate} &= \G{\indMap} \dfis{\indGate-1} + \Z{\indMap} \fis{\indGate-1}, \\
    \dbis{\indGate} &= \GT{\Ngates - \indGate +1} \dbis{\indGate-1} + \ZT{\Ngates-\indGate+1} \bis{\indGate-1}.
\end{align}
\end{subequations}

By utilizing these relations, the full \ac{HVP} of $\costHS$ is computed within the same two-pass budget as the gradient. 
This approach leverages the multi-linear structure of the \ac{TN} to extract second-order curvature without ever explicitly forming or storing the Hessian matrix.

\subsection{Bond dimension of tangent states \label{sec:tangent-states-bond-dimension}}

Whether the \ac{TN} contractions present in the \ac{HVP} computations are efficient depends heavily on the required bond dimensions. 
Specifically, the bond dimension of the tangent states must remain bounded during recursive propagation to ensure computational scalability. 
In what follows, we consider the forward tangent states as an example, but the same argument applies to the backward tangent states.
We denote the maximum bond dimension of the forward states as $\chi$.

Evaluating the recursive expression in \cref{eq:Dpsi} for a forward tangent state $\dfis{\indGate}$ explicitly yields:
\begin{equation}
    \label{eq:dfis-explicit}
    \dfis{\indGate} = \sum_{j=1}^\indGate \br{\pa{\prod_{p=\indGate}^{j+1} \G{p}}  \Z{j} \pa{\prod_{q=j-1}^1 \G{q}}} \fis{0}.
\end{equation}
Because the variation $\Z{j}$ acts on the identical local physical space $\hilbert_{i,i+1}$ as $\G{j}$, each individual term in this summation forms a distinct \ac{MPS} with a maximum virtual bond dimension of $\chi$.
Evaluated naively, this linear combination would result in an additive maximum bond dimension of $\indGate\chi$~\cite{paeckel2019time,schollwock2011density}, leading to an unscalable memory growth with respect to the  number of gates.
We can avoid this by propagating the tangent state together with the unperturbed state in a shared gauge, thereby fixing the bond dimension to $2\chi$ by construction.
This requires a projection at every gate.

\paragraph*{Joint state and augmented gate.}
We represent the unperturbed state and its tangent state as a joint state
\begin{equation}
    \Psi^{[\indGate]} = \begin{pmatrix} \fis{\indGate} \\ \dfis{\indGate} \end{pmatrix} \in \hilbert \oplus \hilbert
\end{equation}
as in forward-mode \ac{AD} with dual numbers~\cite{griewank2008evaluating}.
The recursive propagation of this joint state is governed by the augmented block-triangular operator $\mc{W}^{[\indGate]}$
\begin{equation}
\mc{W}^{[\indGate]} = \begin{pmatrix} \G{\indGate} & 0 \\ \Z{\indGate} & \G{\indGate} \end{pmatrix}, \hspace{1cm} \Psi^{[\indGate]} = \mc{W}^{[\indGate]} \Psi^{[\indGate-1]},
\end{equation}
and multiplying two neighboring augmented gates recovers this block-triangular structure:
\begin{equation*}
\mc{W}^{[\indGate]} \mc{W}^{[\indGate-1]} = \begin{pmatrix} \G{\indGate}\G{\indGate-1} & 0 \\ \Z{\indGate}\G{\indGate-1} + \G{\indGate}\Z{\indGate-1} & \G{\indGate}\G{\indGate-1} \end{pmatrix}.
\end{equation*}
Importantly, the zero-valued top-right block prevents the tangent history from mixing back into the unperturbed state, while the bottom row reproduces exactly the recursion given in \cref{eq:Dpsi}.

\paragraph*{The tangent state in \ac{MPS} form.}
Consider the \ac{MPS} $\fis{\indGate}$ with the local tensors $\{ A_i\}_{i=1}^N$.
Each local tensor lives in $A_i \in \mc{V}\otimes \C^2\otimes\mc{V}$, where $\text{dim}(\mc{V}) = \chi$.
We store the tangent state by means of a second set of local tensors $\{ B_i\}_{i=1}^N$,  $B_i \in \mc{V}\otimes \C^2\otimes\mc{V}$ with the same gauge as $\{ A_i\}_{i=1}^N$:
\begin{equation}
    \dfis{\indGate} = \sum_{i=1}^N \left( \prod_{j=1}^{i-1} A_j \right) B_i \left( \prod_{j'=i+1}^{N} A_{j'} \right).
\end{equation}
Note that this sum runs over all sites, whereas \cref{eq:dfis-explicit} runs over the first $\indGate$ gates.
This is the general form of a tangent vector to the manifold of \acp{MPS} with bond dimension $\chi$~\cite{haegeman2011time,haegeman2016unifying}.
Furthermore, this forms an \ac{MPS} on the augmented virtual space $\mc{V}_\text{aug} = \mc{V}\oplus\mc{V}$, i.e., $\text{dim}(\mc{V}_\text{aug})=2\chi$ as follows:
\begin{align}
    M_i &= \begin{pmatrix}
        A_i & 0 \\
        B_i & A_i\\
    \end{pmatrix}, \\
    \fis{\indGate} &= \begin{pmatrix}
        1 & 0
    \end{pmatrix} 
    \left[\prod_{i=1}^N M_i \right]
    \begin{pmatrix}
        1 \\
        0
    \end{pmatrix}, \\
    \dfis{\indGate} &= \begin{pmatrix}
        0 & 1
    \end{pmatrix} 
    \left[\prod_{i=1}^N M_i \right]
    \begin{pmatrix}
        1 \\
        0
    \end{pmatrix}.
\end{align}
Importantly, the $2\chi$-dimensional tensors $M_i$ are never explicitly built; instead, only the pair $(A_i, B_i)$ is stored.

\paragraph*{Applying an augmented gate.}
Consider gate $\indGate$ acting on the neighboring sites $(i,i+1)$ with the canonical center on one of them.
Applying $\mc{W}^{[\indGate]}$ to $M_i M_{i+1}$ results in an augmented two-site tensor
\begin{align}
    \Theta_\text{aug} = \begin{pmatrix}
        \Theta & 0 \\
        D\Theta & \Theta
    \end{pmatrix},
\end{align}
where $\Theta = \G{\indGate}(A_i A_{i+1}) \in \mc{V}\otimes\hilbert_{i,i+1}\otimes\mc{V}$ is the standard unperturbed 2-site tensor
\begin{align}
    \begin{gathered} \includegraphics[width=4cm]{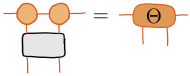} \end{gathered},
\end{align}
and the corresponding tangent two-site tensor is given by 
\begin{equation}
\begin{split}
    D\Theta &= \G{\indGate}(B_i A_{i+1} + A_i B_{i+1}) + \Z{\indGate}(A_i A_{i+1}) \\
    &\in \mc{V}\otimes \mc{H}_{i,i+1}\otimes \mc{V}.
\end{split}
\end{equation}
The tangent local tensors $B_i$ on all other sites are left unchanged.

\paragraph*{Restoring the \ac{MPS} form.}
To restore the block structure of the augmented two-site tensor, we aim for a factorization
\begin{equation}
\begin{split}
    \begin{pmatrix}\Theta&0\\D\Theta&\Theta\end{pmatrix}&\approx
\begin{pmatrix}L&0\\X&L\end{pmatrix}\begin{pmatrix}R&0\\Y&R\end{pmatrix}\\
&=\begin{pmatrix}LR&0\\XR+LY&LR\end{pmatrix}.
\end{split}
\end{equation}
By comparison, we see that $\Theta \approx LR$.
As in unperturbed \ac{TEBD}, this factorization is obtained from the truncated \ac{SVD} of the matricized two-site tensor, $\Theta \approx U_\chi \Sigma_\chi V_\chi^\dagger$, by setting $L=U_\chi$ and $R=\Sigma_\chi V_\chi^\dagger$.

Furthermore, $XR + LY \approx D\Theta$ is required.
We first note that $XR$ preserves the kept right Schmidt vectors and $LY$ the kept left ones.
Their sum can therefore represent any change of $\Theta$ except one that introduces new Schmidt vectors on both sides of the bond simultaneously.
Changes involving kept Schmidt vectors on both sides can be represented by either term; 
we fix this freedom by the gauge condition $U_\chi^\dagger X = 0$~\cite{haegeman2016unifying,lubich2013dynamical}, which yields
\begin{equation}
    Y = U_\chi^\dagger\, D\Theta, \hspace{0.7cm} X=(\Id - P_{U_\chi})\, D\Theta\, V_\chi\Sigma_\chi^{-1},
\end{equation}
with $P_{U_\chi} = U_\chi U_\chi^\dagger$ and $P_{V_\chi} = V_\chi V_\chi^\dagger$.
Thus, $XR + LY$ is the projection of $D\Theta$ onto the tangent space of \acp{MPS} with bond dimension $\chi$: 
\begin{equation}
\begin{split}
    P_T(D\Theta) &= XR + LY \\
    &= P_{U_\chi}\, D\Theta + (\Id - P_{U_\chi})\, D\Theta\, P_{V_\chi},
\end{split}
\end{equation}
which introduces an error that is controlled by the truncated singular values.
The updated local tensors are $(A_i, B_i) \gets (L, X)$ and $(A_{i+1}, B_{i+1}) \gets (R, Y)$, both of bond dimension $\chi$.

After every gate, $A_i$ and $B_i$ have bond dimension of at most $\chi$.
The tangent state is therefore an \ac{MPS} with bond dimension of at most $2\chi$, independent of the number of gates $\indGate$.

The backward tangent states are computed in the same procedure, using the augmented gate built from $\GT{\indGate}$ and $\ZT{\indGate}$.

\subsection{Algorithm to evaluate derivatives of the Hilbert-Schmidt test}

Since the Hilbert-Schmidt test is a direct function of the overlap $\overlap$, its derivatives are efficiently computed by postprocessing the output of \cref{alg:hvp-kernel-core}. 
This integrated procedure is detailed in \cref{alg:HVP-HST}.

\begin{algorithm}[H]
\caption{Computing Hilbert-Schmidt derivatives}\label{alg:HVP-HST}
\begin{algorithmic}[1]
\Require $\psi^{[0]}$, $\phi^{[0]}$, $\vec{G}$, $\vec{V}$
\State $\overlapVar, \grad\overlapVar, \hvp{\overlapVar}{\Zs} \gets$ \Call{HVP\_Kernel}{$\psi^{[0]}, \phi^{[0]}, \vec{G}, \vec{V}$}  \Comment{Call \cref{alg:hvp-kernel-core}}

\Procedure{Postprocess cost function}{$\overlapVar$}
\State $\costHS \gets 1 - \abs{\overlapVar}^2$
\EndProcedure

\Procedure{Postprocess gradient}{$\overlapVar,\grad\overlapVar$}
\State $\grad\costF \gets - \conj{\pa{\grad\overlapVar}}$
\State $\grad \costHS \gets -2 \cdot \overlapVar \cdot \grad \costF$ 
\EndProcedure

\Procedure{Postprocess HVP}{$\overlapVar,\grad\costF, \hvp{\overlapVar}{\Zs}, \Zs$}
\State $\Omega \gets -\sum_{k} \langle \gradi{\indGate} \costF, \Z{\indGate} \rangle$
\State $\hvp{\costF}{\Zs} \gets -\conj{\pa{\hvp{\overlapVar}{\Zs}}}$
\State $\hvp{\costHS}{\Zs} \gets 2 \br{ \Omega \cdot \grad\costF + \overlapVar \cdot \hvp{\costF}{\Zs}}$
\EndProcedure

\State \Return $\costHS, \grad \costHS, \hvp{\costHS}{\Zs}$
\end{algorithmic}
\end{algorithm}

We further extend this framework in \cref{ap:TI} to handle specialized architectures of translationally invariant circuits. 
These structures are particularly important for the study of periodic many-body systems and provide additional symmetries that can be exploited to achieve further computational gains.

%% file: 5_application.tex
\section{Numerical application: quantum circuit compression \label{sec:riemannian-opt}}

We demonstrate the practical utility of our \ac{HVP} kernel through the task of quantum circuit compression -- the process of approximating a deep target unitary evolution with a shallow parameterized ansatz~\cite{robertson2023approximate,zhang2024scalable,kotil2024riemannian,causer2024scalable,gibbs2024deep,le2025riemannian,d2025circuit,misciasci2026exploitingweylchamberstructurehardwareaware,karacan2026nearoptimalquantumtimeevolution}. 
To optimize these circuits while strictly preserving their gate geometry, we employ a Riemannian trust-region algorithm~\cite[Algorithm 10]{absil2008optimization}\cite{kotil2024riemannian} in conjunction with a truncated \ac{CG} method~\cite[Algorithm 11]{absil2008optimization}\cite{kotil2024riemannian} constrained to the unitary manifold.

Historically, the application of second-order optimization in this domain has been restricted to small systems where the full Hessian matrix remains computationally tractable~\cite{kotil2024riemannian}.
As system sizes scale, the memory and time complexities of this explicit construction quickly become prohibitive bottlenecks, such that optimizations have been restricted to first order only~\cite{le2025riemannian,zhang2024scalable}.
Our approach bypasses this limitation entirely: by utilizing \cref{alg:HVP-HST} to evaluate the Hessian's action directly, we avoid the explicit construction of the second-order derivative matrix and thereby extend the reach of Riemannian trust-region methods to significantly larger quantum systems. 

For a comprehensive mathematical treatment of Riemannian optimization, we refer the reader to Ref.~\cite{absil2008optimization}, and to Refs.~\cite{luchnikov2021qgopt,luchnikov2021riemannian,le2025riemannian,kotil2024riemannian} for specialized discussions regarding its application to quantum technologies and quantum circuit compression.

\subsection{Quantum circuit compression as supervised learning}

We formulate quantum circuit compression as a supervised learning task~\cite{zhang2024scalable} by minimizing the Hilbert-Schmidt distance between a reference unitary $\Uref$ -- typically a high-order Trotterization with negligible approximation error -- and a parameterized brickwall ansatz $\brickwall$~\cite{khatri2019quantum, zhang2024scalable, le2025riemannian, gibbs2024deep}:
\begin{equation}
\label{eq:exact-optimization-problem}
\underset{\Gs\in\mc{U}(4)^{\times \Ngates}}{\min} \costHSTot(\Uref, \brickwall).
\end{equation}
In practice, the global objective in \cref{eq:exact-optimization-problem} can be approximated by the \emph{empirical risk}, defined as the average cost over a finite training set of $\Nsamples$ Haar-random \emph{product states} $\fisVarSample\sim\HaarProdSet$~\cite{caro2023out, zhang2024scalable}:
\begin{equation}
\label{eq:costHS}
\begin{aligned}
    \hat{C}_\text{HS}(\Uref,\brickwall) &= 1 - \frac{1}{\Nsamples} \sum_{\indSample=1}^{\Nsamples} \abs{\bisVarSample\brickwall\fisVarSample}^2 \\
    &= 1 - \frac{1}{\Nsamples} \sum_{\indSample = 1}^{\Nsamples} \costHSSample,
\end{aligned}
\end{equation}
where each $\costHSSample$ follows \cref{eq:cost-summands}, and $\bisVarSample = \Uref\fisVarSample$ serves as the corresponding reference label state.
The compression task thus seeks an optimal parameter set $\Gopt$ that maximizes the mean state fidelity across the training ensemble:
\begin{equation}
    \label{eq:approx-optimization-problem}
    \Gopt = \arg\min_{\Gs \in \mathcal{U}(4)^{\times \Ngates}} \hat{C}_{\text{HS}}(\Uref, \brickwall).
\end{equation}

This formulation leverages the powerful in- and out-of-distribution generalization capabilities of variational quantum circuits~\cite{caro2023out, zhang2024scalable}. 
By optimizing the ansatz on a discrete set of product states, the model effectively captures the underlying algebraic structure of $\Uref$. 
This allows the compressed representation to accurately predict the evolution of previously unseen, \emph{general} Haar-random states with high fidelity.

To solve \cref{eq:approx-optimization-problem}, we utilize \cref{alg:HVP-HST} to evaluate the Euclidean gradient and \ac{HVP}. 
However, because the optimization is constrained to the unitary manifold $\mc{U}(4)^{\times \Ngates}$, these Euclidean quantities must be mathematically projected onto the manifold's corresponding tangent spaces. 
We employ Riemannian projections to appropriately constrain these derivatives, which ensures that the trust-region algorithm's search trajectory remains strictly valid; the specific post-processing maps for these projections are detailed in \cref{ap:projection}. 
Furthermore, because the empirical risk is formulated as a sum over independent training samples, the evaluation of the total cost and its derivatives is \emph{embarrassingly parallel}~\cite{10.1145/1146381.1146382}, facilitating highly efficient large-scale execution.

\begin{figure*}[htb]
    \centering
    \begin{subfigure}[t]{0.49\textwidth}
        \centering
        \includegraphics[width=0.9\textwidth]{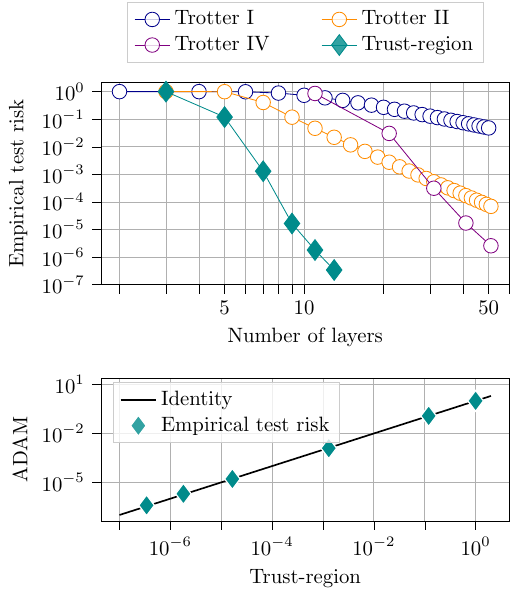}
        \caption{Ising chain with 50 sites and parameters $J=1$, $g=0.75$, and $h=0.6$}
        \label{fig:rqcopt-results-ising}
    \end{subfigure}%
    \hfill
    \begin{subfigure}[t]{0.49\textwidth}
        \centering
        \includegraphics[width=0.9\textwidth]{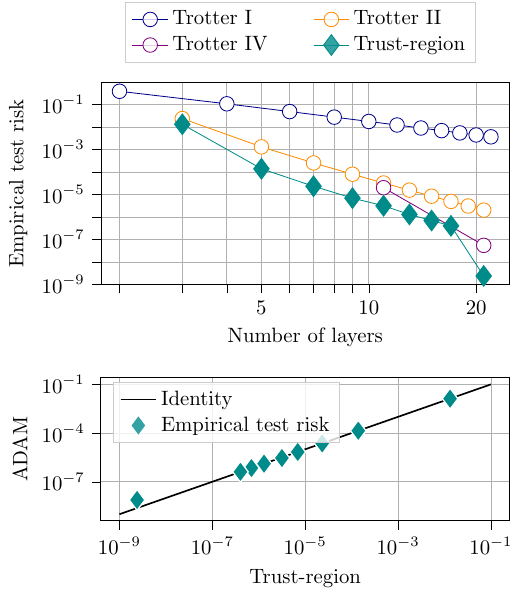}
        \caption{Heisenberg chain with 40 sites and parameters $\vb*{J}=(1,1,-1/2)$ and $\vb*{h}=(3/4,0,0)$.}
        \label{fig:rqcopt-results-heisenberg}
    \end{subfigure}
    \caption{Riemannian quantum circuit optimization for (a) the Ising and (b) the Heisenberg chain. (Top) Empirical test risk, \cref{eq:costHS}, on 100 Haar-random product states as a function of the number of brickwall layers. \emph{Trotter I, II,} and \emph{IV} denote unoptimized first-, second-, and fourth-order Trotter circuits for the same total time $t$. Each layer count corresponds to a different time step. \emph{Trust-region} denotes the optimized circuit obtained by the trust-region method. The reference $\Uref$ is obtained as $20$ steps of Trotter IV. (Bottom) Final test risk of the trust-region method compared with Riemannian ADAM~\cite{le2025riemannian}. Points on the identity line indicate equal accuracy, while points above the identity line indicate smaller loss obtained through the trust-region algorithm.}
    \label{fig:rqcopt-results}
\end{figure*}
\subsection{Numerical results}

To evaluate the performance of the \ac{HVP} kernel, we optimize Trotter circuits using a second-order Riemannian trust-region algorithm~\cite{absil2008optimization}. 
We benchmark its efficacy against the first-order Riemannian ADAM optimizer utilized in our recent study~\cite{le2025riemannian}.
In the following, we will denote the unoptimized first-, second-, and fourth-order Trotter circuits as \emph{Trotter I, II,} and \emph{IV}, respectively. 

We study two spin chains with open boundary conditions.
The first is a transverse-field Ising model on a chain of $N=50$ sites:
\[
H^\text{Ising} = \sum_{i=1}^{N-1}  J Z_i Z_{i+1} + \sum_{i=1}^{N} \left( gX_i + hZ_i \right),
\]
where $J,g,h\in\R$. 
We set $J=1$, $g=0.75$, and $h=0.6$, for which the model is non-integrable, and simulate up to time $t=2$.
The second is a Heisenberg chain of $N=40$ sites: 
\[
H^\text{Heis} = \sum_{i=1}^{N-1} \sum_{\alpha=1}^3 J^\alpha \sigma_i^\alpha \sigma_{i+1}^\alpha + \sum_{i=1}^N \sum_{\alpha=1}^3 h^\alpha \sigma_i^\alpha,
\]
where $\vec{J},\vec{h}\in \R^3$. 
For this model, we parameterize the system with $\vec{J}=(1,1,-1/2)$ and $\vec{h}=(3/4,0,0)$, and consider a total simulation time of $t=0.25$. 
For both models, the reference unitary $\Uref$ is given by Trotter IV with $20$ time steps.
Throughout, we use a maximum bond dimension of $\chi=38$ for the Ising model, and $\chi=32$ for the Heisenberg model.

We initialize the brickwall ansatz with Trotter II and Trotter IV circuits for the total time $t$, using different numbers of Trotter steps. 
Since a finer time discretization results in a deeper circuit, each initialization corresponds to a different number of brickwall layers. 
These circuits are then optimized over a training set of $N_s = 16$ samples for the Ising model and $N_s=8$ samples for the Heisenberg model. 
Detailed documentation of the optimized models and their corresponding optimization parameters is provided in \cref{ap:optimization-models}.
All accuracies are quantified by the empirical risk (\cref{eq:costHS}), evaluated on a test batch of 100 previously unseen Haar-random product states, and are shown in \cref{fig:rqcopt-results} (top panel) as a function of the number of layers.

\begin{figure*}[htb]
    \centering
    \begin{subfigure}[t]{0.49\textwidth}
        \centering
        \includegraphics[width=0.9\textwidth]{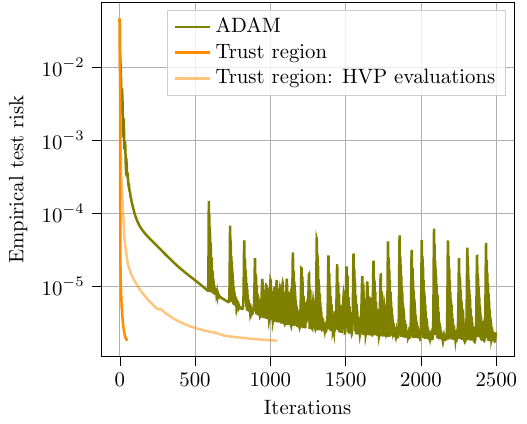}
        \caption{Ising chain for 50 sites with parameters $J=1$, $g=0.75$, and $h=0.6$. The trust-region algorithm requires 1044 \ac{HVP} evaluations ($T_\text{wall}=4.85$\,h), while the Riemannian optimizer requires 2500 gradient evaluations ($T_\text{wall}=8.64$\,h).}
        \label{fig:convergence-comparison-ising}
    \end{subfigure}%
    \hfill
    \begin{subfigure}[t]{0.49\textwidth}
        \centering
        \includegraphics[width=0.9\textwidth]{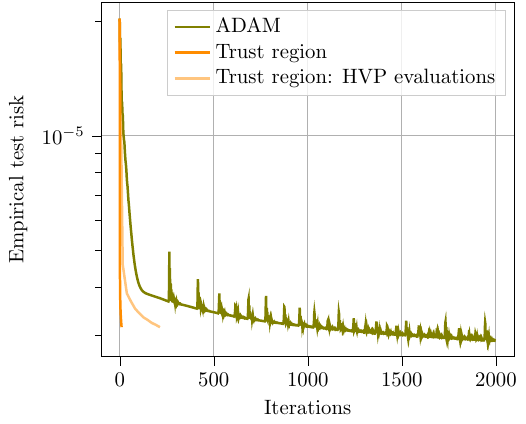}
        \caption{Heisenberg chain for 40 sites with parameters $\vb*{J}=(1,1,-1/2)$ and $\vb*{h}=(3/4,0,0)$. The trust-region algorithm requires 214 \ac{HVP} evaluations ($T_\text{wall}=15$\,min) in comparison to 1999 gradient evaluations when using the Riemannian ADAM ($T_\text{wall}= 160$\,min).}
        \label{fig:convergence-comparison-heisenberg}
    \end{subfigure}
    \caption{Convergence of Riemannian ADAM and the trust-region method for (a)~the Ising and (b)~the Heisenberg chain, for a translationally invariant brickwall circuit with 11 layers. The empirical test loss is evaluated on 100 Haar-random product states. Solid lines show the loss as a function of the number of optimization steps. The dotted orange line shows the same trust-region run against the cumulative number of \ac{HVP} evaluations, since each trust-region step requires several \ac{HVP} evaluations within the truncated CG solver. For Riemannian ADAM, each step requires exactly one gradient evaluation, so the number of steps and the number of derivative evaluations coincide. ADAM step sizes are $\eta = 0.01$ (Ising) and $\eta = 0.001$ (Heisenberg).}
    \label{fig:convergence-comparison}
\end{figure*}

The bottom panel of \cref{fig:rqcopt-results} shows that our second-order approach either reproduces or improves upon the results of Ref.~\cite{le2025riemannian}.
Compared with the initial Trotter circuits, the optimized circuits reduce the approximation error by up to four orders of magnitude for the Ising model and by up to one order of magnitude for the Heisenberg model.
For circuits with up to 17 layers, the trust-region method and Riemannian ADAM reach comparable final errors.
However, we observe that in these cases, the trust-region algorithm exhibits better convergence.
We show this as an example for quantum circuits composed of 11 layers in \cref{fig:convergence-comparison}.
Because Riemannian ADAM performs a single gradient evaluation per update, whereas each trust-region step utilizes the truncated \ac{CG} method~\cite[Algorithm 11]{absil2008optimization}, which involves multiple \ac{HVP} evaluations, comparing them solely by iteration count is insufficient. 
To ensure a fair comparison, we report the loss as a function of both the total number of iterations and the cumulative number of derivative evaluations.
For the Ising chain, the trust-region method requires 1044 \ac{HVP} evaluations, compared with 2500 gradient evaluations for ADAM, reducing the total optimization time from $8.64$\,h to $4.85$\,h. 
For the Heisenberg chain, the corresponding numbers are 214 and 1999, and the optimization time decreases from $160$\,min to $15$\,min, meaning a wall-time reduction by one order of magnitude.
In \cref{sec:spectral}, we analyze the spectrum of the Riemannian Hessian to argue why we observe this rapid convergence.
The Riemannian ADAM optimizer exhibits a somewhat unstable convergence profile, characterized by frequent fluctuations and spikes in the loss function. 
This instability is a direct consequence of it relying purely on first-order information; lacking knowledge of the local curvature, the optimizer frequently overshoots minima in high-curvature or ill-conditioned regions of the manifold. 
This effect can be ameliorated by decreasing the step size, at the cost of more iterations.
In contrast, the second-order trust-region algorithm yields significantly smoother and monotonic convergence. 
By leveraging our \ac{HVP} kernel, the algorithm incorporates the local Hessian, thereby adaptively constraining the step size within a ``trusted region'' where the quadratic approximation of the landscape remains valid. 
This geometric awareness prevents the overshooting characteristic of first-order methods, resulting in a significantly lower iteration count. 
However, we note that a reduction in iterations does not guarantee shorter runtime in general, because one \ac{HVP} evaluation costs about three to four times as much as one gradient evaluation.
We therefore give a complete documentation of the evaluation times in \cref{ap:optimization-models}.

\begin{figure}[htb]
\color{blue}
\centering
    \vspace{0pt}
    \captionsetup{width=\linewidth, margin=0pt}
    \centering
    \includegraphics[width=0.9\linewidth]{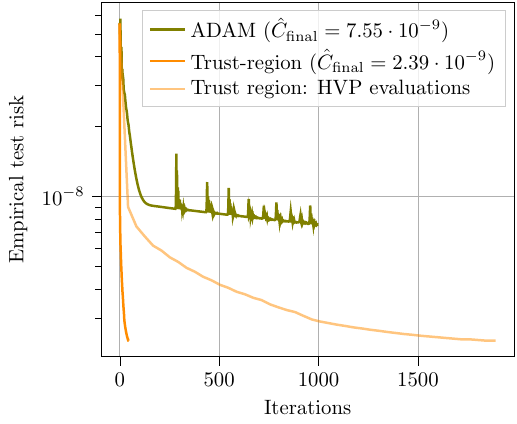}
    \caption{Loss convergence curve for a quantum circuit of $21$ layers. The Riemannian ADAM optimizer gets stuck in a local minimum ($\hat{C}_\text{final} = 7.55\cdot 10^{-9}$), whereas the trust-region algorithm reduces the loss by more than half ($\hat{C}_\text{final} = 2.39\cdot 10^{-9}$).}
    \label{fig:heisenberg_IV_2}
\end{figure}

\begin{figure*}[ht]
    \centering
    \includegraphics[width=0.75\linewidth]{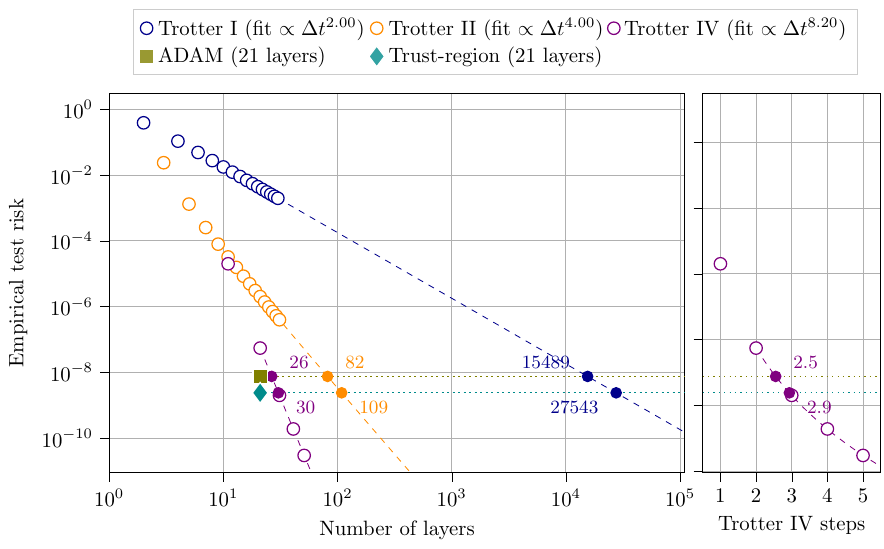}
    \caption{Compression of a Trotter IV$_2$ circuit (21 layers) by Riemannian ADAM and the trust-region method. Dashed lines are power-law fits to the Trotter errors, used to extrapolate Trotter I and II beyond the computed data.
    (Left) Layers that Trotter I, II, and IV circuits need to match the test risk of the optimized circuits; these need not be valid Trotter depths (e.g., Trotter II has $2n_\text{steps}+1$ layers).
    (Right) Equivalent number of Trotter IV steps. The trust-region method with our \ac{HVP} kernel nearly compresses three Trotter IV steps into two; Riemannian ADAM, using only gradients, stagnates between two and three steps and thus does not achieve this compression.
    }
    \label{fig:heisenberg-savings}
\end{figure*}

For deeper circuits, the trust-region method can reach significantly lower errors.
For the Heisenberg model initialized with two Trotter IV steps, its final error is less than half that of ADAM.
The convergence curves in \cref{fig:heisenberg_IV_2} show that Riemannian ADAM stagnates early, presumably in a local minimum, whereas the trust-region method continues to decrease the loss but requires many \ac{HVP} evaluations to do so.
This indicates a more difficult loss landscape for deeper circuits, in which curvature information helps the optimizer reach better solutions.
To quantify the practical benefit, we compare the optimized circuits with Trotter IV circuits, which, for this model, achieve a given accuracy with the fewest layers among the product formulas considered (\cref{fig:heisenberg-savings}).
Both optimizations are initialized with two Trotter IV steps, i.e., 21 layers.
Since the number of Trotter steps is an integer, the relevant question is whether an optimized 21-layer circuit reaches the accuracy of the next deeper Trotter IV circuit, i.e., three repetitions (31 layers).
The trust-region method achieves this as shown in \cref{fig:heisenberg-savings}.
Three Trotter IV steps are thus compressed into the depth of two, reducing the circuit depth by about one third.
Riemannian ADAM also improves on the initial circuit, but stagnates at $7.55\cdot10^{-9}$, between the errors of two and three Trotter IV steps, and therefore does not reach the accuracy of any deeper Trotter IV circuit.

%% file: 6_conclusion.tex
\section{Conclusion}

In this work, we introduced an analytical framework for evaluating \acp{HVP} tailored to \acp{TN}. 
Based on recursive tangent-state propagation, the proposed \ac{HVP} kernel provides access to global second-order curvature information without the prohibitive memory and computational overhead of explicitly constructing the full Hessian. 
We guarantee the scalability of this approach by demonstrating that the maximum virtual bond dimension of the tangent states remains strictly bounded, ensuring computational feasibility even for more complex architectures.
Quasi-Newton methods such as (L-)BFGS, which have Riemannian extensions~\cite{doi:10.1137/140955483,doi:10.1137/11082885X}, also avoid forming the Hessian, but only build a positive-definite approximation of it from gradient history. In contrast, the exact \ac{HVP} gives the true curvature, including negative-curvature directions, which truncated CG within a trust region can exploit near saddle points. The two approaches are complementary, and the \ac{HVP} kernel could equally serve inside Newton–CG schemes or be combined with quasi-Newton preconditioners.

To demonstrate its practical utility, we integrated the proposed \ac{HVP} kernel into a Riemannian trust-region framework applied to quantum circuit compression. 
Evaluated on non-integrable transverse-field Ising and Heisenberg spin chains, our second-order optimizer reproduced previous results of first-order quantum circuit compressions for shallow circuits.
Crucially, incorporating geometric curvature enabled the trust-region method to achieve significantly smoother and more monotonic convergence than first-order methods such as Riemannian ADAM, 
without tuning a step size. For the models we considered, we obtained a runtime reduction of up to one order of magnitude.
Furthermore, for deep quantum circuits, using our \ac{HVP} kernel in a second-order optimization resulted in improved final solutions.
This enabled us to compress three Trotter IV steps into two, which is not possible with only first-order derivatives.
By adaptively constraining the step size, our framework prevents the erratic overshooting characteristic of first-order methods in high-curvature or ill-conditioned regions of the manifold, providing a more direct and stable path to the optimum.
This interpretation is further supported by the eigenvalue spectrum of the Riemannian Hessian, which reveals the strong ill-conditioning underlying the instability of first-order methods.

The scalability of our \ac{HVP} framework opens several promising avenues for future research. 
A natural extension is to investigate its utility for learning circuits with infinite \acp{TN}~\cite{gibbs2025learning} in the thermodynamic limit. 
Beyond 1D architectures, the trust-region optimization approach could be used for second-order optimization in variational Monte Carlo with projected entangled pair states~\cite{sandvik2007variational, liu2017gradient,liu2021accurate,wu2026algorithms,chen2026variational}, analogous to such efforts with neural quantum states~\cite{webber2022rayleigh}. 
By integrating this kernel into 2D many-body simulations, it may be possible to navigate the complex energy landscapes of frustrated systems with greater stability than traditional stochastic reconfiguration methods. 
Furthermore, this framework offers a scalable lens into the second-order geometry of large-scale systems; combining the \ac{HVP} with Lanczos-based methods~\cite{lanczos1950iteration} would allow for the efficient evaluation of Hessian spectra and condition numbers, providing a powerful diagnostic tool for characterizing the loss landscape of \acp{TN}.

%% file: 7_acknowledgement.tex
\section*{Acknowledgments}
The Riemannian ADAM optimizer is based on the ADAM implementation in \texttt{qiskit v0.18}~\cite{qiskit2024}, while the trust-region method is adapted from \texttt{rqcopt v1.0.0}~\cite{rqcopt2022github}. 
All underlying tensor network operations were executed using \texttt{jax.numpy}~\cite{jax2018github}. 
The authors gratefully acknowledge the computational and data resources provided by the Leibniz Supercomputing Centre (\url{www.lrz.de}).
I.L.\ and R.W.\ thank Lukas Devos for insightful discussions regarding automatic differentiation and second-order optimization methods. 
I.L.\ further thanks Lukas Brenner for valuable conversations on reverse-over-reverse computations. 
The Flatiron Institute is a division of the Simons Foundation. 

\subsection*{Data availability}
The implementation used in this project, the numerical results, and a script to generate the figures in this publication are available at \url{https://github.com/INMLe/rqcopt-hvp}.

%% file: acronyms.tex
\begin{acronym}
    \acro{MPO}{matrix product operator}
    \acro{MPS}{matrix product state}
    \acro{TEBD}{time-evolving block-decimation}
    \acro{HVP}{Hessian-vector product}
    \acro{TN}{tensor network}
    \acro{AD}{automatic differentiation}
    \acro{SVD}{singular value decomposition}
    \acro{CG}{conjugate gradient}
\end{acronym}

\lstset{
  basicstyle=\ttfamily,
  columns=fullflexible,
  frame=single,
}

%% file: 8_appendix.tex
\setcounter{page}{1}
\renewcommand\thefigure{\thesection\arabic{figure}}
\setcounter{figure}{0} 
\onecolumngrid

\begin{center}
\large{ Supplementary Material for: \\``Scalable Hessian-vector products for tensor networks with application to quantum circuit compression''
}
\end{center}

\section{Complex derivatives \label{ap:wirtinger}}

In this section, we summarize the framework for complex differentiation utilized throughout this work, specifically the Wirtinger calculus~\cite{koor2023short}.

Let $f:\C\rightarrow\C$ be a (not necessarily holomorphic) smooth function, and $z=x+iy\in\C$ with $x,y\in\R$. 
The Wirtinger derivatives are defined as:
\begin{align}
    \pd{f}{z} &= \frac{1}{2} \pa{\pd{f}{x} - i\pd{f}{y}}, & \pd{f}{\conj{z}} &= \frac{1}{2} \pa{\pd{f}{x} + i\pd{f}{y}}.
\end{align}
The complex gradient of $f$ is given by:
\begin{equation}
    \grad f(z) = \pd{f}{x} + i \pd{f}{y} = 2\cdot \conj{\pa{\pd{f}{z}}} = 2\cdot \pd{f}{\conj{z}}.
\end{equation}
We also note the following conjugate relations:
\begin{subequations}
\begin{align}
    \pd{\conj{f}}{\conj{z}} &= \frac{1}{2} \pa{\pd{\conj{f}}{x} + i\pd{\conj{f}}{y}} = \conj{\br{\frac{1}{2} \pa{\pd{f}{x} - i\pd{f}{y}}}} = \conj{\br{ \pd{f}{z}}}, \\
    \pd{\conj{f}}{z} &= \frac{1}{2} \pa{\pd{\conj{f}}{x} - i\pd{\conj{f}}{y}} = \conj{\br{ \frac{1}{2} \pa{\pd{f}{x} + i\pd{f}{y}}}} = \conj{\br{\pd{f}{\conj{z}}}}.
\end{align}
\end{subequations}
The latter derivative vanishes for any holomorphic function $f$. 

For a matrix-valued function $f:\C^{m\times m}\rightarrow \C$, these definitions apply entry-wise:
\begin{equation}
    \pa{\pd{f}{G}}_{ij} = \pd{f}{G_{ij}}.
\end{equation}
If $f$ depends on a sequence of $\Ngates$ matrices, $\Gs = (\G{1},\ldots,\G{\Ngates})$, the complex gradient of $f$, denoted as $\grad f(\Gs) = \br{\gradi{\indGate} f(\Gs)}_{\indGate = 1 \ldots \Ngates}$, is constructed component-wise:
\begin{equation}
    \gradi{\indGate} f(\Gs) = 2\cdot \conj{\pa{\pd{f}{\G{\indGate}}}}.
\end{equation}

The Wirtinger product rule states:
\begin{equation}
    \pd{(fg)}{z} = \pa{\pd{f}{z}} g + f \pa{\pd{g}{z}}.
\end{equation}
Using this, the gradient of the product \( f(z)g(z) \) is computed as:
\begin{equation} 
    \nabla(fg) = 2 \left[ \left( \frac{\partial f}{\partial z} \right) g + f \left( \frac{\partial g}{\partial z} \right) \right]^* = 2 \left( \frac{\partial f}{\partial z} \right)^\ast g^\ast + 2 f^\ast \left( \frac{\partial g}{\partial z} \right)^\ast = (\nabla f) \cdot g^\ast + f^\ast \cdot (\nabla g).
\end{equation}

For a composition $h = f \circ g$ with a complex-valued function $g:\C\rightarrow\C$, the Wirtinger chain rule reads:
\begin{align}
    \pd{h}{z} &= \pd{f}{g} \pd{g}{z} + \pd{f}{\conj{g}} \pd{\conj{g}}{z}, & \pd{h}{\conj{z}} &= \pd{f}{g} \pd{g}{\conj{z}} + \pd{f}{\conj{g}} \pd{\conj{g}}{\conj{z}}.
\end{align}
Consequently, the complex gradient of the composition is:
\begin{equation}
    \grad h = 2 \pa{\pd{f}{g} \pd{g}{\conj{z}} + \pd{f}{\conj{g}} \pd{\conj{g}}{\conj{z}}} = 2 \pa{\pd{f}{g} \pd{g}{\conj{z}} + \pd{f}{\conj{g}} \conj{\pa{\pd{g}{z}}}}.
\end{equation}

The first-order variation of a function $f:\C\rightarrow\C$ along a direction $v\in\C$ is captured by the directional derivative (with $\epsilon\in\R$):
\begin{align*}
    \dgrad{f(z)}{v} = \lim_{\epsilon \to 0} \frac{f(z + \epsilon v) - f(z)}{\epsilon}.
\end{align*}
This definition extends naturally to vector-valued functions $\vb*{f}:\C^d\rightarrow\C^d$.

Given another function $g:\C\rightarrow\C$, the directional derivative of the product $fg$ is derived as follows:
\begin{equation}
\label{ap:eq:dgrad-product}
\begin{aligned}
D\big((fg)(z)\big)[v]
    &= \lim_{\varepsilon\to 0}\frac{f(z+\varepsilon v)\,g(z+\varepsilon v)-f(z)\,g(z)}{\varepsilon}\\
    &= \lim_{\varepsilon\to 0}\frac{f(z+\varepsilon v)\,g(z+\varepsilon v)-f(z)\,g(z+\varepsilon v)+f(z)\,g(z+\varepsilon v)-f(z)\,g(z)}{\varepsilon}\\
    &= \lim_{\varepsilon\to 0}\left[\frac{f(z+\varepsilon v)-f(z)}{\varepsilon}\,g(z+\varepsilon v)+f(z)\,\frac{g(z+\varepsilon v)-g(z)}{\varepsilon}\right]\\
    &= D\big(f(z)\big)[v]\,g(z)+f(z)\,D\big(g(z)\big)[v].
\end{aligned}
\end{equation}

Similarly, the product rule for the outer product of two vector-valued functions $\vb*{f},\vb*{g}:\C^d\rightarrow\C^d$ yields:
\begin{equation}
\label{ap:eq:dgrad-product-vec}
    D\big((\vb*{f}\otimes\vb*{g})(\vb*{z})\big)[\vb*{v}]
= D\big(\vb*{f}(\vb*{z})\big)[\vb*{v}]\otimes\vb*{g}(\vb*{z})
+ \vb*{f}(\vb*{z})\otimes D\big(\vb*{g}(\vb*{z})\big)[\vb*{v}].
\end{equation}

\section{Computation of $\Omega$ \label{ap:Omega}}

To compute the directional derivative of the total overlap $\overlapVar(\Gs)$, we first isolate the contribution of a single gate $\G{\indGate}$. We denote this localized overlap as:
\begin{equation}
    \overlapVar^{[\indGate]}(\G{\indGate}) = \sum_{ij\alpha} \bisC{\Ngates-\indGate}_{i\alpha} \G{\indGate}_{ij} \fis{\indGate-1}_{j\alpha}. 
\end{equation}

Introducing a perturbation $\epsilon \Z{\indGate}$ to this isolated gate yields the first-order expansion:
\begin{equation}
\begin{split}
    \overlapVar^{[\indGate]} \pa{ \G{\indGate}+\epsilon \Z{\indGate} } &= \sum_{ij\alpha} \bisC{\Ngates-\indGate}_{i\alpha} \br{ \G{\indGate} + \epsilon \Z{\indGate} }_{ij} \fis{\indGate-1}_{j\alpha} 
    \\
    &= \sum_{ij\alpha} \bisC{\Ngates-\indGate}_{i\alpha} \G{\indGate}_{ij} \fis{\indGate-1}_{j\alpha} + \epsilon \sum_{ij\alpha} \bisC{\Ngates-\indGate}_{i\alpha} \Z{\indGate}_{ij}  \fis{\indGate-1}_{j\alpha}.
\end{split}
\end{equation}

The directional derivative of the global overlap along a collective variation direction $\Zs$ is defined as the sum of these local perturbations evaluated at $\epsilon=0$. Grouping the appropriate terms, the derivation proceeds as follows:
\begin{equation}
\begin{split}
    \Omega \equiv \dgrad{\overlap}{\Zs} &= \sum_{\indGate = 1}^\Ngates \lim_{\epsilon\to0} \frac{\overlapVar^{[\indGate]}(\G{\indGate} + \epsilon \Z{\indGate})}{\epsilon} 
    = \sum_{\indGate=1}^\Ngates \sum_{ij\alpha} \bisC{\Ngates-\indGate}_{i\alpha} \Z{\indGate}_{ij} \fis{\indGate-1}_{j\alpha} \\
    &= \sum_{\indGate=1}^\Ngates \sum_{ij} \pa{ \sum_{\alpha} \bisC{\Ngates-\indGate}_{i\alpha} \fis{\indGate-1}_{j\alpha} } \Z{\indGate}_{ij}
    = \sum_{\indGate=1}^\Ngates \sum_{ij} \pa{\pd{\overlap}{\G{\indGate}}}_{ij} \Z{\indGate}_{ij} \\
    &= \sum_{\indGate=1}^\Ngates \Tr\br{\pa{-\gradi{\indGate}\costF(\Gs)}^\dagger \Z{\indGate}} 
    = -\sum_{\indGate=1}^{\Ngates} \langle \gradi{\indGate}\costF(\Gs), \Z{\indGate} \rangle,
\end{split}
\end{equation}
where we substituted the complex gradient $\grad\costF$ as defined in \cref{eq:grad-costF} and utilized the Hilbert-Schmidt inner product $\langle A, B\rangle = \Tr(A^\dagger B)$. Note that, unlike the real-valued cost functions, this directional derivative evaluates to a complex scalar ($\Omega\in\C$) in general.

\section{Deriving the HVP as the gradient of the directional derivative \label{ap:RoR-derivation}}   

In this section, we demonstrate that an analytic evaluation of the \ac{HVP} via the gradient of the directional derivative,
\begin{align*}
    \hvp{f(\As)}{\Zs}  = \grad \br{\dgrad{f(\As)}{\Zs}},
\end{align*} 
(cf.\ \cref{eq:RoR} in the main text) simply yields \cref{alg:hvp-kernel-core}.

To streamline the subsequent derivations, we adopt a slightly modified notation compared to the main text. 
Specifically, we denote the inner product between two quantum states $\fisVar, \bisVar \in\hilbert$ as
\[
    \TT{\bisVar} \fisVar = \sum_i \bisVar_i \fisVar_i.
\]
By absorbing the complex conjugation directly into the dual state, we only need to manipulate states in the Hilbert space and its complex conjugate (rather than explicitly tracking the dual space), which significantly simplifies the algebraic bookkeeping.

The \ac{HVP} is computed as the gradient of the directional derivative. The directional derivative of the overlap $\overlap$ is given by (cf.\ \cref{eq:Omega} in the main text):
\begin{equation*}
    \Omega \equiv \dgrad{\overlapVar(\As)}{\Zs} = \sum_{\indMap=1}^{\Nmaps} \omega^{[\indMap]},
\end{equation*}
where the summand 
\begin{equation}
    \label{ap:eq:omega}
    \omega^{[\indMap]} = \bisTT{\Nmaps-\indMap} \Z{\indMap} \fis{\indMap-1} = \sum_{ij} \bis{\Nmaps-\indMap}_{i} \Z{\indMap}_{ij} \fis{\indMap-1}_{j}
\end{equation}
corresponds to the overlap wherein the $\indMap$-th local map is replaced by its corresponding perturbation $\Z{\indMap}$. 

Because $\Omega$ remains a holomorphic function with respect to the parameters $\As$, its complex gradient is simply defined by the partial derivatives $\partial\Omega/\partial \A{\indMap}$. Hence, we arrive at the expression:
\begin{equation}
    \label{ap:eq:overlap-hvp-RoR}
    \hvp{\overlapVar(\As)}{\Zs} = \grad \Omega = \pa{\pd{\Omega}{\A{\indMap}}}_{\indMap=1\ldots \Nmaps}.
\end{equation}

As we will show, this formulation naturally recovers the exact same forward and backward passes derived via the forward-over-reverse mode. While both approaches are mathematically equivalent due to the symmetry of mixed partial derivatives, they differ in memory and computational requirements when evaluated using black-box \ac{AD}. 

Isolating a single variational local map $\A{\indMap}$, the partial derivative of $\Omega$ is obtained via the chain rule:
\begin{equation}
    \label{ap:eq:dOmega}
    \pd{\Omega}{\A{\indMap}} = \underbrace{\pd{\Omega}{\fis{\indMap}}}_{\bfis{\indMap}} \otimes \pd{\fis{\indMap}}{\A{\indMap}} + \underbrace{\pd{\Omega}{\bis{\Nmaps-\indMap+1}}}_{\bbis{\Nmaps-\indMap+1}} \otimes \pd{\bis{\Nmaps-\indMap+1}}{\A{\indMap}},
\end{equation}
where we have identified and introduced the \textit{dual tangent states}:
\begin{equation}
    \bfis{\indMap} = \pd{\Omega}{\fis{\indMap}} \qquad \text{and} \qquad \bbis{\Nmaps-\indMap+1} = \pd{\Omega}{\bis{\Nmaps-\indMap+1}}.
\end{equation}

\subsubsection*{Computing the dual tangent states $\bfis{\indMap}$}

At the boundary, we evaluate the initial conditions:
\begin{align}
    \bfis{\Nmaps} &= \pd{\Omega}{\fis{\Nmaps}} = 0, & \bfis{\Nmaps-1} &= \pd{\Omega}{\fis{\Nmaps-1}} = \underbrace{\pd{\Omega}{\omega^{[\Nmaps]}}}_{=1} \cdot \pd{\omega^{[\Nmaps]}}{\fis{\Nmaps-1}}.
\end{align}
For the remaining states $\bfis{\indMap-1}$, expanding via the chain rule yields:
\begin{equation}
    \bfis{\indMap-1} = \pd{\Omega}{\fis{\indMap-1}} = \underbrace{\pd{\Omega}{\omega^{[\indMap]}}}_{=1} \cdot \pd{\omega^{[\indMap]}}{\fis{\indMap-1}} + \underbrace{\pd{\Omega}{\fis{\indMap}}}_{\bfis{\indMap}} \cdot \pd{\fis{\indMap}}{\fis{\indMap-1}} = \pd{\omega^{[\indMap]}}{\fis{\indMap-1}} + \bfis{\indMap} \pd{\fis{\indMap}}{\fis{\indMap-1}}.
\end{equation}
Since $\fis{\indMap} = \A{\indMap}\fis{\indMap-1}$, the partial derivative evaluates to $\pd{\fis{\indMap}}{\fis{\indMap-1}} = \A{\indMap}$. Expressing the second term component-wise, we find:
\begin{equation}
    \br{\bfis{\indMap} \cdot \pd{\fis{\indMap}}{\fis{\indMap-1}}}_{m} = \sum_{i} \bfis{\indMap}_{i} \cdot \pd{\fis{\indMap}_{i}}{\fis{\indMap-1}_{m}} = \sum_i \bfis{\indMap}_{i} \A{\indMap}_{im} = \br{\ATT{\indMap} \bfis{\indMap} }_{m}.
\end{equation}
Furthermore, evaluating the first term using \cref{ap:eq:omega} gives:
\begin{equation}
    \br{\pd{\omega^{[\indMap]}}{\fis{\indMap-1}}}_{m} =  \pd{}{\fis{\indMap-1}_{m}} \br{ \sum_{ij} \bis{\Nmaps-\indMap}_{i} \Z{\indMap}_{ij} \fis{\indMap-1}_{j}} = \sum_{i} \bis{\Nmaps-\indMap}_{i} \Z{\indMap}_{im} = \br{\ZTT{\indMap} \bis{\Nmaps-\indMap}}_{m}.
\end{equation}
Combining these components, we obtain the final backward recursion:
\begin{align}
    \bfis{\Nmaps} &= 0, & \bfis{\indMap-1} &= \ZTT{\indMap} \bis{\Nmaps-\indMap} + \ATT{\indMap} \bfis{\indMap}.
\end{align}

\subsubsection*{Computing the dual tangent states $\bbis{\Nmaps-\indMap+1}$}

Similarly, to compute the backward dual tangent states, we establish the boundary conditions:
\begin{align}
    \bbis{\Nmaps} &= 0, & \bbis{\Nmaps-1} &= \pd{\Omega}{\bis{\Nmaps-1}} = \underbrace{\pd{\Omega}{\omega^{[1]}}}_{=1} \cdot \pd{\omega^{[1]}}{\bis{\Nmaps-1}}.
\end{align}
For the intermediate states $\bbis{\Nmaps-\indMap}$, the chain rule expands to:
\begin{align}
    \bbis{\Nmaps-\indMap} &= \pd{\Omega}{\bis{\Nmaps-\indMap}} = \underbrace{\pd{\Omega}{\omega^{[\indMap]}}}_{=1} \cdot \pd{\omega^{[\indMap]}}{\bis{\Nmaps-\indMap}} + \underbrace{\pd{\Omega}{\bis{\Nmaps-\indMap+1}}}_{=\bbis{\Nmaps-\indMap+1}} \cdot \pd{\bis{\Nmaps-\indMap+1}}{\bis{\Nmaps-\indMap}} = \pd{\omega^{[\indMap]}}{ \bis{\Nmaps-\indMap}} + \bbis{\Nmaps-\indMap+1} \pd{ \bis{\Nmaps-\indMap+1}}{\bis{\Nmaps-\indMap}}.
\end{align}
Given the propagation rule $\bis{\Nmaps-\indMap+1} = \ATT{\indMap}\bis{\Nmaps-\indMap}$, we have $\pd{\bis{\Nmaps-\indMap+1}}{\bis{\Nmaps-\indMap}} = \ATT{\indMap}$. The second term thus evaluates to:
\begin{equation}
    \br{ \bbis{\Nmaps-\indMap+1} \cdot \pd{\bis{\Nmaps-\indMap+1}}{\bis{\Nmaps-\indMap}} }_{m} = \sum_{i} \bbis{\Nmaps-\indMap+1}_{i} \pd{\bis{\Nmaps-\indMap+1}_{i}}{\bis{\Nmaps-\indMap}_{m}} = \sum_{i} \bbis{\Nmaps-\indMap+1}_{i} \A{\indMap}_{mi} = \br{\A{\indMap} \bbis{\Nmaps-\indMap+1} }_{m}.
\end{equation}
Applying \cref{ap:eq:omega} to the first term yields:
\begin{equation}
    \br{\pd{\omega^{[\indMap]}}{\bis{\Nmaps-\indMap}}}_{m} = \pd{}{\bis{\Nmaps-\indMap}_{m}} \sum_{ij} \bis{\Nmaps-\indMap}_{i} \Z{\indMap}_{ij} \fis{\indMap-1}_{j} = \sum_{j} \Z{\indMap}_{mj} \fis{\indMap-1}_{j} = \br{\Z{\indMap}\fis{\indMap-1}}_{m}.
\end{equation}
Consolidating these results provides the forward recursion:
\begin{align}
    \bbis{\Nmaps} &= 0, & \bbis{\Nmaps-\indMap} &= \Z{\indMap}\fis{\indMap-1} + \A{\indMap}\bbis{\Nmaps-\indMap+1}.
\end{align}

\subsubsection*{Equivalence between primary and dual tangent states}

Crucially, the dual states $\bfis{\indMap-1}$ can be computed recursively via a backward pass, whereas the states $\bbis{\Nmaps-\indMap}$ are computed recursively via a forward pass.
To demonstrate that these quantities coincide exactly with the primary tangent states defined in \cref{eq:dfis,eq:dbis} of the main text, we must align their indexing conventions. 
In the preceding derivation, the dual states were initialized at the boundary conditions $\bfis{\Nmaps}$ and $\bbis{\Nmaps}$. 
By formally reversing this numbering scheme such that the boundary conditions are instead denoted as $\bfis{0}$ and $\bbis{0}$, we obtain the re-indexed recurrence relations:
\begin{subequations}
\begin{align}
    \bfis{0}&=0, &\bfis{\indMap} &= \ZTT{\Nmaps-\indMap+1} \bis{\indMap-1} + \ATT{\Nmaps-\indMap+1} \bfis{\indMap-1}, \\
    \bbis{0}&=0, &\bbis{\indMap} &= \Z{\indMap}\fis{\indMap-1} + \A{\indMap}\bbis{\indMap-1}.
\end{align}
\end{subequations}

Reconciling the inner product convention by restoring the standard adjoint operator ($\ATT{\indMap} \rightarrow \AT{\indMap}$), we recover the exact identities:
\begin{equation}
    \bbis{\indMap} = \dfis{\indMap} \qquad \text{and} \qquad \bfis{\indMap} = \dbis{\indMap}.
\end{equation}

Conceptually, this establishes a powerful symmetry: taking the partial derivative of $\Omega$ (the directional derivative of the global overlap) with respect to a forward (backward) intermediate state is mathematically equivalent to computing the directional derivative of the complementary backward (forward) intermediate state.

\subsubsection*{Computing the Hessian-vector product}

To conclude the derivation, we evaluate the terms of \cref{ap:eq:dOmega}. 
We first observe that the initial term evaluates directly to an outer product:
\begin{equation}
    \bfis{\indMap} \cdot \pd{\fis{\indMap}}{\A{\indMap}} = \bfis{\indMap} \otimes \fis{\indMap-1}.
\end{equation}

For the second term, we compute the partial derivative component-wise with respect to the matrix elements $\A{\indMap}_{mn}$:
\begin{equation}
    \br{\bbis{\Nmaps-\indMap+1} \cdot \pd{\bis{\Nmaps-\indMap+1}}{\A{\indMap}}}_{mn} = \sum_{i} \bbis{\Nmaps-\indMap+1}_{i} \pd{}{\A{\indMap}_{mn}} \pa{ \sum_j \A{\indMap}_{ji} \bis{\Nmaps-\indMap}_{j}} = \bbis{\Nmaps-\indMap+1}_{n} \bis{\Nmaps-\indMap}_{m}.
\end{equation}
Reconstructing this into matrix notation yields the second outer product:
\begin{equation}
    \bbis{\Nmaps-\indMap+1} \cdot \pd{\bis{\Nmaps-\indMap+1}}{\A{\indMap}} = \bis{\Nmaps-\indMap} \otimes \bbis{\Nmaps-\indMap+1}.
\end{equation}

Consolidating these results, we arrive at the expression for the $\indMap$-th component of the global \ac{HVP}:
\begin{equation}
\begin{aligned}
    \br{\hvp{\overlapVar(\As)}{\Zs}}_{\indMap} &= \bfis{\indMap} \otimes \fis{\indMap-1} + \bis{\Nmaps-\indMap} \otimes \bbis{\Nmaps-\indMap+1}\\
    &= \dbis{\Nmaps-\indMap} \otimes \fis{\indMap-1} + \bis{\Nmaps-\indMap} \otimes \dfis{\indMap-1}.
\end{aligned}
\end{equation}
This formulation demonstrates that the local curvature at step $\indMap$ is naturally expressed as a superposition of two outer products, establishing exact equivalence with \cref{eq:overlap-hvp} in the main text.

\section{Modified HVP kernel for translationally invariant brickwall circuits \label{ap:TI}}

In this section, we assume that each layer $\indLayer$ in the brickwall circuit consists of the same layer gate $\Gl{\indLayer}$.
We denote the set of all gate indices $\indGate$ that belong to a layer $\indLayer$ as $\indLayerList$.
In other words, for each gate $\G{\indGate}$ that is located in layer $\indLayer$, it is: $\G{\indGate} = \Gl{\indLayer}$.
This means that the vector of gates $\Gs$ that form the brickwall circuit now consists of only $\Nlayers$ gates. The same thing holds for $\Zs$.

\subsubsection*{Computing the complex gradient}

The gradient is now given with respect to each layer gate $\Gl{\indLayer}$.
It is,
\begin{equation}
\begin{split}
    \pd{\overlap}{\Gl{\indLayer}} &= \sum_{\indGate\in\indLayerList} \pd{\overlap}{\G{\indGate}} = \sum_{\indGate\in\indLayerList} \pa{\sum_\alpha \bis{\Ngates-\indGate}_{i\alpha} \fis{\indGate-1}_{j\alpha}}_{ij} \\
    &= \begin{minipage}[h]{8cm}
	\vspace{0pt}
	\includegraphics[width=8cm]{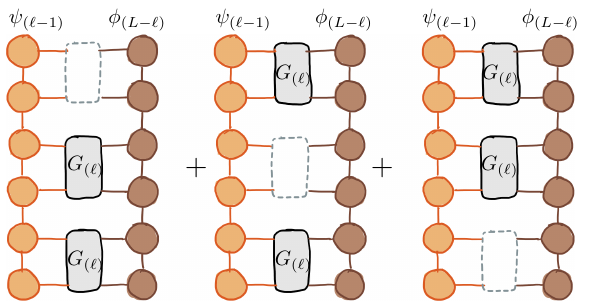}
   \end{minipage}
\end{split}
\end{equation}
In the \ac{TN} diagram, $\fisVar_{(\indLayer)}$ and $\bisVar_{(\indLayer)}$ denote the intermediate states before layer $\indLayer$ is applied.
The complex gradient of the Frobenius norm and Hilbert-Schmidt test are built from this Wirtinger derivative as given in \cref{eq:grad-costF,eq:grad-costHS} in the main text.

\subsubsection*{Computing the directional derivative of the overlap}

In the translationally invariant case, we assume the perturbation $\Zs$ is also layer-wise constant. 
That is, for every gate position $\indGate$ in layer $\indLayer$, the direction is the same: $\Z{\indGate}=\Zl{\indLayer}$.
The directional derivative of the overlap can then be computed as 
\begin{equation}
\begin{split}
    \Omega &= \sum_{\indLayer=1}^\Nlayers \sum_{\indGate\in\indLayerList} \Tr\br{\ZlT{\indLayer} \pd{\overlap}{\G{\indGate}}} = \sum_{\indLayer=1}^\Nlayers \Tr\br{ \ZlT{\indLayer} \pa{\sum_{\indGate\in\indLayerList}  \pd{\overlap}{\G{\indGate}}} } \\
    &= \sum_{\indLayer=1}^\Nlayers \Tr\br{ \ZlT{\indLayer} \pd{\overlap}{\Gl{\indLayer}}} = -\sum_{\indLayer=1}^\Nlayers \langle \gradi{\indLayer} \costF, \Zl{\indLayer} \rangle
\end{split}
\end{equation}

\subsubsection*{Computing the Hessian-vector product of the overlap}

We are interested in the \ac{HVP} of the overlap for the case of translational invariance.
This can be computed as 
\begin{align}
    \hvp{\overlap}{\Zs} = \dgrad{\grad\costF}{\Zs} =\br{\dgrad{\pd{\overlap}{\Gl{\indLayer}}}{\Zs}}_{\indLayer=1\ldots\Nlayers} = \br{ \sum_{\indGate\in\indLayerList} \dgrad{\pd{\overlap}{\G{\indGate}}}{\Zs}}_{\indLayer=1 \ldots \Nlayers}.
\end{align}
The \ac{HVP} of $\costF$ and $\costHS$ can be built from the \ac{HVP} of the overlap as given in \cref{eq:hvp-costF,eq:hvp-costHS}.

A modified algorithm to compute the derivatives of the overlap is presented in \cref{alg:HVP-TI}.

\begin{figure*}
\begin{minipage}{\linewidth}
\begin{algorithm}[H]
\caption{Analytical \ac{HVP} kernel for translationally invariant systems}\label{alg:HVP-TI}
\begin{algorithmic}[1]
\Require $\fis{0}, \bis{0}, \Gs, \Zs$
\Procedure{HVP\_Kernel\_TI}{$\fis{0}, \bis{0}, \Gs, \Zs$}
    \Procedure{Forward pass}{$\fis{0}, \Gs, \Zs$}
        \For{$\indLayer = 1, \ldots, \Nlayers$}
            \For{$\indGate \in \indLayerList$} \Comment{Sum over contributions in layer}
                \State $\dfis{\indGate} \gets \Gl{\indLayer} \dfis{\indGate-1} + \Zl{\indLayer} \fis{\indGate-1}$ \Comment{Compute and cache the corresponding directional derivative}
                \State $\fis{\indGate} \gets \Gl{\indLayer}\fis{\indGate-1}$ \Comment{Compute and cache the forward intermediate state}
            \EndFor
        \EndFor
    \EndProcedure
    \Procedure{Backward pass}{$\bis{0}, \{ \fis{\indGate} \}_{\indGate=1\ldots \Ngates}, \{\dfis{\indGate} \}_{\indGate=1\ldots \Ngates}, \Gs, \Zs$}
        \State $\bisVar \gets \bis{0}$
        \State $\dbisVar \gets 0$
        \For{$\indLayer = \Nlayers, \ldots, 1$}
            \State $\gradi{\indLayer}\overlapVar \gets 0$
            \State $\dgrad{\pd{\overlapVar}{\Gl{\indLayer}}}{\Zs} \gets 0$
            \For{$\indGate \in\indLayerList $}
                \State $\dbisVar \gets \GlT{\indLayer} \dbisVar + \ZlT{\indLayer} \bisVar$ \Comment{Update the directional derivative of the intermediate state}
                \State $\gradi{\indLayer}\overlapVar \gets \gradi{\indLayer}\overlapVar-\sum_\alpha \bisVar_{i\alpha} \fis{\Ngates-\indGate-1}_{j\alpha}$  \Comment{Build up the complex gradient}
                \State $\hvp{\overlapVar}{\Zs} \gets \hvp{\overlapVar}{\Zs} + \sum_\alpha \dbisVar_{i\alpha} \fis{\Ngates-\indGate}_{j\alpha} + \sum_\alpha \bisVar_{i\alpha} \dfis{\Ngates-\indGate}_{j\alpha}$ \Comment{Build up the \ac{HVP}}
                \State $\bisVar \gets \GlT{\indLayer}\bisVar$ \Comment{Update the backward intermediate state}
            \EndFor
        \EndFor
        \State $\overlap \gets \bis{0}\fis{\Ngates}$
    \EndProcedure
\EndProcedure
\Ensure $\overlapVar, \grad\overlapVar, \hvp{\overlapVar}{\Zs}$
\end{algorithmic}
\end{algorithm}
\end{minipage}
\end{figure*}

\section{Projecting Euclidean derivatives to the unitary tangent bundle \label{ap:projection}}

We consider the unitary manifold of a single two-qubit gate, $\mc{U}(4)$, equipped with the Riemannian metric induced by the Hilbert-Schmidt inner product $\langle A,B\rangle = \text{Tr}(A^{\dagger}B)$. 
Any complex matrix $V\in\C^{4\times 4}$ is projected onto the tangent space $\mc{T}_G\mc{U}(4)$ of point $G\in\mc{U}(4)$ via the orthogonal projection:
\begin{align}\label{eq:projec}
    P_G(V) &= G\,\text{skew}\pa{G^\dagger V} = \frac{1}{2} V - \frac{1}{2} G V^\dagger G.
\end{align}

For the full circuit, the parameter set $\Gs=(\G{1},\G{2},\dots,\G{\Ngates})$ resides on the product manifold $\mc{U}(4)^{\times \Ngates}$. 
Consequently, the global Riemannian gradient $\grad_R f$ is obtained by applying the localized projection $P_{\G{\indGate}}$ to the Euclidean derivatives of each gate $\indGate \in \{1,\dots,\Ngates\}$ independently:
\begin{align}
\grad_R f = \br{P_{\G{\indGate}}\pa{\grad_{\indGate} f}}_{\indGate = 1,\dots,\Ngates}.
\end{align}

The Riemannian \ac{HVP} can be obtained by first taking the directional derivative of the Riemannian gradient and then projecting this result back to the tangent space. 
To this end, we first consider only one parameter, i.e., one gate $G$ and a corresponding perturbation $V$. 
We take the directional derivative of the Riemannian gradient as follows:
\begin{equation}
\begin{aligned}
    \dgrad{\grad_R f}{V} &= \dgrad{ P_G \pa{\grad f}}{V} = \frac{1}{2} \dgrad{\grad f - G(\grad f)^\dagger G}{V} \\
    &= \frac{1}{2} \dgrad{ \grad f}{V} - \frac{1}{2} \dgrad{G(\grad f)^\dagger G }{V} = \frac{1}{2} \hvp{f}{V} - \frac{1}{2} \dgrad{G(\grad f)^\dagger G }{V}.
\end{aligned}
\end{equation}
The second term can be obtained by the product rule as
\begin{equation}
\begin{aligned}
    \dgrad{G(\grad f)^\dagger G }{V} &= V(\grad f)^\dagger G + G\,\dgrad{\grad f^\dagger}{V} G  + G(\grad f)^\dagger V  \\
    &= V(\grad f)^\dagger G + G\,\pa{\dgrad{\grad f}{V}}^\dagger G  + G(\grad f)^\dagger V \\
    &= V(\grad f)^\dagger G + G\,\pa{\hvp{f}{V}}^\dagger G  + G(\grad f)^\dagger V
\end{aligned}
\end{equation}
and with this, we arrive at
\begin{equation}
\begin{aligned}
    \label{eq:directional-deriv-of-riemannian-gradient}
     \dgrad{ \grad_R f }{V} &= \frac{1}{2} \hvp{f}{V} - \frac{1}{2} V(\grad f)^\dagger G - \frac{1}{2} G\,\pa{\hvp{f}{V}}^\dagger G  - \frac{1}{2} G(\grad f)^\dagger V \\
     &= P_G\pa{ \hvp{f}{V} } - \frac{1}{2} V(\grad f)^\dagger G - \frac{1}{2} G(\grad f)^\dagger V.
\end{aligned}
\end{equation}
To obtain the Riemannian \ac{HVP}, we need to apply another projection on \cref{eq:directional-deriv-of-riemannian-gradient}. 
We note that $P_G^2 = P_G$ and with that 
\begin{align}
    H_R(f)[V] &= P_G\pa{\dgrad{\grad_R f}{V}} = P_G\pa{\hvp{f}{V} - \frac{1}{2} V \grad f^\dagger G - \frac{1}{2} G\grad f^\dagger V }
\end{align}

To apply this to a quantum circuit consisting of multiple gates $\Gs$, these projections can be applied gate-wise:
\begin{align}
H_R(f)[\Z{\indGate}] &= \br{P_{\G{\indGate}} \pa{\hvp{f}{\Z{\indGate}} - \frac{1}{2} \Z{\indGate} (\grad_{\indGate} f)^\dagger \G{\indGate} - \frac{1}{2} \G{\indGate} (\grad_\indGate f)^\dagger \Z{\indGate} }}_{\indGate = 1, \dots, \Ngates}.
\end{align}

\section{Optimization models and execution times of derivatives \label{ap:optimization-models}}

We study two spin chains with open boundary conditions.
The first is a transverse-field Ising model on a chain of $N=50$ sites:
\[
H^\text{Ising} = \sum_{i=1}^{N-1}  J Z_i Z_{i+1} + \sum_{i=1}^{N} \left( gX_i + hZ_i \right),
\]
where $J,g,h\in\R$. 
The second is a Heisenberg chain of $N=40$ sites: 
\[
H^\text{Heis} = \sum_{i=1}^{N-1} \sum_{\alpha=1}^3 J^\alpha \sigma_i^\alpha \sigma_{i+1}^\alpha + \sum_{i=1}^N \sum_{\alpha=1}^3 h^\alpha \sigma_i^\alpha,
\]
where $\vec{J},\vec{h}\in \R^3$. 
For both models, the reference unitary $\Uref$ is given by Trotter IV with $20$ time steps.
In the following, we report the models that were optimized, along with their optimization setups.
We further document the overall wall time and the peak memory usage of the optimizations.
However, we want to clarify that no early-stopping criteria were used, and the chosen number of iterations therefore determines the total execution duration.

\begin{table}[h]
\centering
\caption{Optimized brickwall circuits for the Ising model ($L=50$, $t=2$, $J=1$, $g=0.75$, $h=0.6$, 16 training and 96 test samples, maximum bond dimension $\chi_\text{max}=38$). $\eta$ is the ADAM learning rate, $T$ the total wall time of the optimization and $M$ its peak memory.}
\label{tab:opt_ising}
\begin{tabular}{cc ccrr crr}
\toprule
 & & \multicolumn{4}{c}{ADAM} & \multicolumn{3}{c}{Trust-region} \\
\cmidrule(lr){3-6} \cmidrule(lr){7-9}
Model & $n_\text{layers}$ & $n_\text{iter}$ & $\eta$ & $T$ [min] & $M$ [GiB] & $n_\text{iter}$ & $T$ [min] & $M$ [GiB] \\
\midrule
II$_1$ & 3  & 500  & $5\cdot 10^{-2}$ & 35.5  & 6.04 & 20 & 8.7   & 4.96 \\
II$_2$ & 5  & 500  & $5\cdot 10^{-2}$ & 49.6  & 5.46 & 20 & 19.7  & 4.92 \\
II$_3$ & 7  & 1000 & $5\cdot 10^{-2}$ & 130.9 & 6.83 & 20 & 41.7  & 4.69 \\
II$_4$ & 9  & 2500 & $1\cdot 10^{-2}$ & 482.5 & 8.17 & 75 & 303.7 & 4.86 \\
II$_5$ & 11 & 2500 & $1\cdot 10^{-2}$ & 518.6 & 8.35 & 50 & 290.7 & 5.10 \\
II$_6$ & 13 & 2500 & $1\cdot 10^{-2}$ & 593.4 & 8.78 & 50 & 351.7 & 5.61 \\
\bottomrule
\end{tabular}
\end{table}

\begin{table}[h]
\centering
\caption{Optimized brickwall circuits for the Heisenberg model ($L=40$, $t=0.25$, $J=(1,1,-0.5)$, $h=(0.75,0,0)$, 8 training and 96 test samples, maximum bond dimension $\chi_\text{max}=32$). $\eta$ is the ADAM learning rate, $T$ the total wall time of the optimization and $M$ its peak memory.}
\label{tab:opt_heisenberg}
\begin{tabular}{cc ccrr crr}
\toprule
 & & \multicolumn{4}{c}{ADAM} & \multicolumn{3}{c}{Trust-region} \\
\cmidrule(lr){3-6} \cmidrule(lr){7-9}
Model & $n_\text{layers}$ & $n_\text{iter}$ & $\eta$ & $T$ [min] & $M$ [GiB] & $n_\text{iter}$ & $T$ [min] & $M$ [GiB] \\
\midrule
II$_1$ & 3  & 100  & $10^{-2}$        & 3.0   & 3.06 & 10 & 1.6   & 2.66 \\
II$_2$ & 5  & 200  & $10^{-2}$        & 8.1   & 3.26 & 10 & 3.3   & 2.82 \\
II$_3$ & 7  & 500  & $5\cdot 10^{-3}$ & 25.3  & 3.56 & 10 & 6.2   & 2.90 \\
II$_4$ & 9  & 1000 & $10^{-3}$        & 63.3  & 4.32 & 10 & 10.1  & 3.01 \\
IV$_1$ & 11 & 2000 & $10^{-3}$        & 160.2 & 5.81 & 10 & 14.8  & 2.77 \\
II$_6$ & 13 & 1000 & $5\cdot 10^{-4}$ & 102.0 & 4.45 & 10 & 23.0  & 2.93 \\
II$_7$ & 15 & 1000 & $5\cdot 10^{-4}$ & 115.9 & 4.46 & 10 & 29.5  & 2.79 \\
II$_8$ & 17 & 1000 & $5\cdot 10^{-4}$ & 135.7 & 4.46 & 10 & 37.4  & 2.93 \\
IV$_2$ & 21 & 1000 & $5\cdot 10^{-5}$ & 144.4 & 4.34 & 45 & 230.5 & 3.35 \\
\bottomrule
\end{tabular}
\end{table}

We further report execution times to evaluate the gradient and \ac{HVP} for the models optimized in \cref{sec:riemannian-opt}.
To this end, we report the mean execution time and the corresponding standard deviation for 10 evaluations each.
The models denoted by II$_n$ correspond to an initial second-order Trotter circuit of $n$ steps; models denoted by IV$_n$ correspond to initial fourth-order Trotter circuits of $n$ steps.

\begin{table}[htb]
\centering
\caption{Execution time in seconds (mean $\pm$ standard deviation over 10 runs, without compilation) and peak memory in GiB of the Riemannian gradient and the Riemannian \ac{HVP} for the one-dimensional Ising model with $L=50$ sites ($16$ samples), bond dimension $\chi=32$. The columns \ac{HVP}/Gradient give the ratio of the mean execution times and of the peak memory, respectively.}
\label{tab:timings_ising}
\begin{tabular*}{\textwidth}{@{\extracolsep{\fill}}lcccccc}
\toprule
 & \multicolumn{3}{c}{Execution time [s]} & \multicolumn{3}{c}{Peak memory [GiB]} \\
\cmidrule(lr){2-4} \cmidrule(lr){5-7}
Model & Gradient & HVP & HVP/Gradient & Gradient & HVP & HVP/Gradient \\
\midrule
$\mathrm{II}_{1}$ & $0.68 \pm 0.01$ & $2.08 \pm 0.03$ & $3.04$ & $0.29$ & $0.53$ & $1.82$ \\
$\mathrm{II}_{2}$ & $1.09 \pm 0.02$ & $3.34 \pm 0.06$ & $3.06$ & $0.36$ & $0.68$ & $1.86$ \\
$\mathrm{II}_{3}$ & $1.52 \pm 0.02$ & $4.59 \pm 0.07$ & $3.02$ & $0.44$ & $0.82$ & $1.88$ \\
$\mathrm{II}_{4}$ & $1.90 \pm 0.02$ & $5.77 \pm 0.10$ & $3.04$ & $0.51$ & $0.96$ & $1.90$ \\
$\mathrm{II}_{5}$ & $3.53 \pm 0.03$ & $10.17 \pm 0.11$ & $2.88$ & $0.58$ & $1.11$ & $1.91$ \\
$\mathrm{II}_{6}$ & $4.19 \pm 0.03$ & $12.05 \pm 0.16$ & $2.87$ & $0.65$ & $1.25$ & $1.92$ \\
\bottomrule
\end{tabular*}
\end{table}

\begin{table}[htb]
\centering
\caption{Execution time in seconds (mean $\pm$ standard deviation over 10 runs, without compilation) and peak memory in GiB of the Riemannian gradient and the Riemannian \ac{HVP} for the Heisenberg model with $L=40$ sites ($8$ samples), bond dimension $\chi=32$. The columns \ac{HVP}/Gradient give the ratio of the mean execution times and of the peak memory, respectively.}
\label{tab:timings_heisenberg}
\begin{tabular*}{\textwidth}{@{\extracolsep{\fill}}lcccccc}
\toprule
 & \multicolumn{3}{c}{Execution time [s]} & \multicolumn{3}{c}{Peak memory [GiB]} \\
\cmidrule(lr){2-4} \cmidrule(lr){5-7}
Model & Gradient & HVP & HVP/Gradient & Gradient & HVP & HVP/Gradient \\
\midrule
$\mathrm{II}_{1}$ & $0.37 \pm 0.01$ & $1.31 \pm 0.01$ & $3.52$ & $0.12$ & $0.21$ & $1.83$ \\
$\mathrm{II}_{2}$ & $0.59 \pm 0.01$ & $2.12 \pm 0.02$ & $3.58$ & $0.15$ & $0.27$ & $1.86$ \\
$\mathrm{II}_{3}$ & $0.73 \pm 0.01$ & $2.49 \pm 0.02$ & $3.42$ & $0.17$ & $0.33$ & $1.88$ \\
$\mathrm{II}_{4}$ & $0.83 \pm 0.01$ & $2.94 \pm 0.04$ & $3.53$ & $0.20$ & $0.39$ & $1.90$ \\
$\mathrm{IV}_{1}$ & $0.98 \pm 0.01$ & $3.51 \pm 0.01$ & $3.57$ & $0.23$ & $0.44$ & $1.91$ \\
$\mathrm{II}_{6}$ & $1.17 \pm 0.01$ & $4.14 \pm 0.04$ & $3.53$ & $0.26$ & $0.50$ & $1.92$ \\
$\mathrm{II}_{7}$ & $1.32 \pm 0.01$ & $4.77 \pm 0.05$ & $3.61$ & $0.29$ & $0.56$ & $1.93$ \\
$\mathrm{II}_{8}$ & $1.48 \pm 0.01$ & $5.30 \pm 0.03$ & $3.58$ & $0.32$ & $0.62$ & $1.94$ \\
$\mathrm{IV}_{2}$ & $1.83 \pm 0.01$ & $6.51 \pm 0.02$ & $3.56$ & $0.37$ & $0.73$ & $1.95$ \\
\bottomrule
\end{tabular*}
\end{table}

\section{Spectral analysis of the Riemannian Hessian for truncated conjugate gradient descent}\label{sec:spectral}

To understand the convergence behavior of the second-order Riemannian optimization, we can construct the dense Riemannian Hessian matrix $\mathbf{\vb*{H}}_R$. This operator is self-adjoint with respect to the
Riemannian metric $\Re\langle A,B\rangle = \text{Tr}(A^{\dagger}B)$
\begin{align*}
    \langle\mathbf{\vb*{H}}_R u,v\rangle = \langle u,\mathbf{\vb*{H}}_R v\rangle.
\end{align*}
and therefore has a spectrum of real eigenvalues. Since $\mathbf{\vb*{H}}_R$ is a linear map between tangent spaces, we can choose a basis $\{E_i\}$ for the tangent space and expand the operator in these coordinates:
\begin{align}
    (\mathbf{\vb*{H}}_R)_{ij} 
    = \langle E_i, \mathbf{\vb*{H}}_R E_j \rangle 
    = \langle E_i,\sum_{k} E_k H_R(f)[E_j]_k \rangle
    =\sum_{k} E_k H_R(f)[E_j]_k  \langle E_i, E_k \rangle
\end{align}
Note that if the basis is orthonormal, $\langle E_i, E_j \rangle=\delta_{ij}$, then the HVP can be used directly to construct the column entries of the matrix.

For a single unitary gate $\G{\indGate}\in\Gs$, we can parameterize the directions in the tangent space by choosing a basis for the Lie algebra $\mathfrak{u}(4)$, which is given by tensor products of the Pauli Operator and the identity. We can then move to the tangent space at $\G{\indGate}\in\Gs$ by left multiplication of $\G{\indGate}$
\begin{align*}
    \mc{B}_{\G{\indGate}}=\cb{i\G{\indGate}(P_a \otimes P_b)/2 \, | \, P_a, P_b \in \{I,X,Y,Z\}}.
\end{align*}
For $K$ gates, this results in a basis set $\{B_i\}$, $i=1,\ldots,\Ngates \cdot 16$. The number of elements in this basis equals $16\Ngates$ since for each gate we have $|\mc{B}_{\G{\indGate}}|=16$ Pauli operators. An example tangent basis vector acting on the second gate is given by $(0, i\G{\indGate}(X\otimes Y)/2, 0,\hdots,0)$, hence we have a single Pauli operator acting on gate $\indGate$, and the other directions are padded with zeros. 

\begin{figure}[hbt]
    \centering
    \begin{subfigure}[t]{0.49\textwidth}
        \centering
        \includegraphics[width=0.9\textwidth]{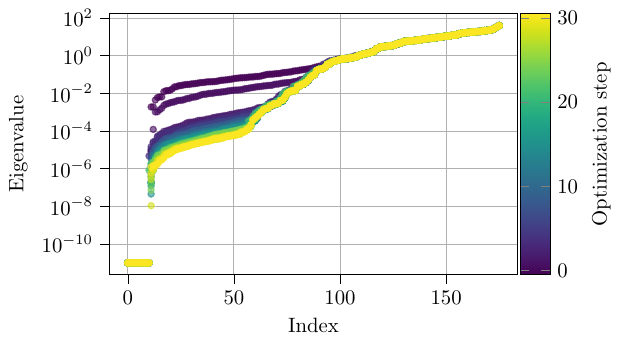}
        \caption{Ising chain for 20 sites with parameters $J=1$, $g=0.75$, and $h=0.6$.}
        \label{fig:spectrum-ising}
    \end{subfigure}%
    \hfill
    \begin{subfigure}[t]{0.49\textwidth}
        \centering
        \includegraphics[width=0.9\textwidth]{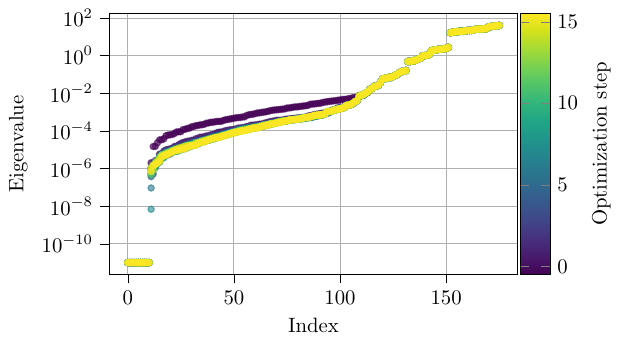}
        \caption{Heisenberg chain for 20 sites with parameters $\vb*{J}=(1,1,-1/2)$ and $\vb*{h}=(3/4,0,0)$.}
        \label{fig:spectrum-heisenberg}
    \end{subfigure}
    \caption{Spectrum of the Riemannian Hessian during an optimization of a brickwall circuit with 11 layers.}
    \label{fig:spectrum}
\end{figure}

Once we have materialized the full Hessian matrix, we can perform an eigenvalue decomposition and analyze the spectrum. As shown in \cref{fig:spectrum}, we see that the resulting spectrum contains eigenvalues spread over multiple orders of magnitude, indicating a high condition number of the Hessian matrix, which can cause problems with iterative linear solvers~\cite{kelley1999iterative}. 

\begin{figure}[hbt]
    \centering
    \begin{subfigure}[t]{0.49\textwidth}
        \centering
        \includegraphics[width=0.9\textwidth]{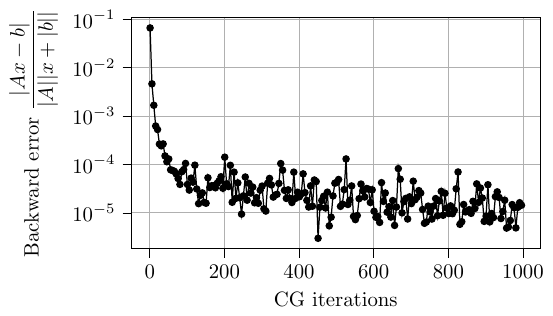}
        \caption{Ising chain for 20 sites with parameters $J=1$, $g=0.75$, and $h=0.6$.}
        \label{fig:cg-ising}
    \end{subfigure}%
    \hfill
    \begin{subfigure}[t]{0.49\textwidth}
        \centering
        \includegraphics[width=0.9\textwidth]{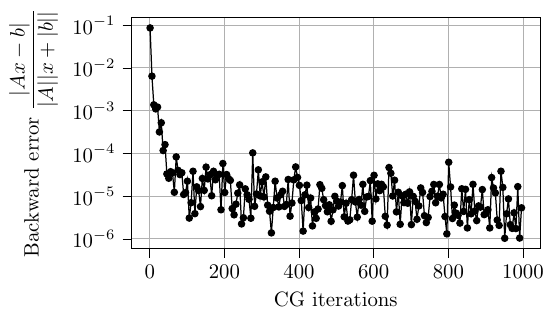}
        \caption{Heisenberg chain for 20 sites with parameters $\vb*{J}=(1,1,-1/2)$ and $\vb*{h}=(3/4,0,0)$.}
        \label{fig:cg-heisenberg}
    \end{subfigure}
    \caption{\Ac{CG} descent convergence for a brickwall circuit with 11 layers.}
    \label{fig:cg}
\end{figure}

However, the spectrum is far from uniformly spread. 
A small number of eigenvalues are large, while the majority lie several orders of magnitude below them, close to zero. 
In CG, the iterates first resolve the components along the dominant eigenvectors~\cite{shewchuk1994cg}. 
Because these large-curvature directions dominate the decrease of the quadratic model, a good approximate solution is obtained after a few iterations, even though full convergence would require resolving the near-flat directions as well. 
This is consistent with \cref{fig:cg}, where the backward error decreases rapidly at first and then stagnates. 
It explains the efficacy of the truncated CG used in the trust-region method: the trust-region step only requires an approximate solution, which is reached within a few CG iterations.

